\documentclass[a4paper,fleqn,usenatbib]{mnras}

\usepackage{newtxtext,newtxmath}

\usepackage[T1]{fontenc}
\usepackage{ae,aecompl}

\usepackage{caption}
\usepackage{booktabs}

\usepackage{color}

\usepackage{graphicx}	
\usepackage{amsmath}	
\usepackage{amssymb}	
\usepackage{siunitx}

\newcommand{\be}{\begin{equation}}
\newcommand{\ee}{\end{equation}} 
\newcommand{\bal}{\begin{align}}
\newcommand{\eal}{\end{align}}
\newcommand{\bnum}{\begin{enumerate}}
\newcommand{\enum}{\end{enumerate}}

\newcommand{\ben}{\begin{enumerate}}
\newcommand{\bi}{\begin{itemize}}
\newcommand{\ei}{\end{itemize}}
\newcommand{\een}{\end{enumerate}}

\newcommand{\om}{\ensuremath{\Omega_\mr m}}

\newcommand{\sig}{\ensuremath{\sigma_8}}
\newcommand{\lcdm}{$\Lambda$CDM}

\newcommand{\mr}{\mathrm}

\newcommand{\halofit}{{\textsc{halofit}}}
\newcommand{\redmagic}{{\textsc{redMaGiC}}}
\newcommand{\zred}{\ensuremath{z_{\textsc{rM}}}}
\newcommand{\zm}{z_{\mathrm{mean}}}
\newcommand{\camb}{{\textsc{camb}}}

\newcommand{\xip}{\ensuremath{\xi_{+}}}
\newcommand{\xim}{\ensuremath{\xi_{-}}}
\newcommand{\xipm}{\ensuremath{\xi_{\pm}}}
\newcommand{\gammat}{\ensuremath{\gamma_{t}(\theta)}}
\newcommand{\wtheta}{\ensuremath{w(\theta)}}

\newcommand{\dx}[1]{d{#1}\,}

\newcommand\eqn[1]{equation~\ref{#1}}
\newcommand\eqnb[2]{equations~\ref{#1}~\& \ref{#2}}

\newcommand\fig[1]{Figure~\ref{#1}}

\newcommand\sect[1]{Section~\ref{#1}}
\newcommand\tab[1]{Table~\ref{#1}}
\newcommand\app[1]{Appendix~\ref{#1}}

\newcommand{\photoz}{photo-$z$}

\newcommand{\ztrue}{\ensuremath{z_{\mathrm{true}}}}

\newcommand{\msun}{\ensuremath{M_{\odot}}}

\DeclareSIUnit \h {\mbox{$h$}}
\DeclareSIUnit \hinv {\mbox{$h^{-1}$}}
\DeclareSIUnit \deg {\mbox{deg}}
\DeclareSIUnit \msun {\mbox{$M_{\odot}$}}

\DeclareSIUnit \parsec {pc}
\DeclareSIUnit \kilaparsec {kpc}
\DeclareSIUnit \megaparsec {Mpc}
\DeclareSIUnit \gigaparsec {Gpc}

\newcommand{\vect}[1]{\boldsymbol{\mathbf{#1}}}

\newcommand{\hinv}{\ensuremath{h^{-1}}}

\usepackage{eso-pic}

\AddToShipoutPictureBG*{%
  \AtPageUpperLeft{%
    \hspace{0.75\paperwidth}%
    \raisebox{-3.5\baselineskip}{%
      \makebox[0pt][l]{\textnormal{DES-2017-0285}}
}}}%

\AddToShipoutPictureBG*{%
  \AtPageUpperLeft{%
    \hspace{0.75\paperwidth}%
    \raisebox{-4.5\baselineskip}{%
      \makebox[0pt][l]{\textnormal{FERMILAB-PUB-18-080}}
}}}%

\title[DES Y1 Results: Validating cosmological parameter estimation using simulated Dark Energy Surveys]{DES Y1 Results: Validating cosmological parameter estimation using simulated Dark Energy Surveys}

\makeatletter
\def \blfootnote{\xdef\@thefnmark{}\@footnotetext}
\makeatother

\author[DES Collaboration]{
\parbox{\textwidth}{
\Large
N.~MacCrann$^{1,2\star}$,
J.~DeRose$^{3,4\dag}$,
R.~H.~Wechsler$^{3,4,5}$,
J.~Blazek$^{1,6}$,
E.~Gaztanaga$^{7,8}$,
M.~Crocce$^{7,8}$,
E.~S.~Rykoff$^{4,5}$,
M.~R.~Becker$^{3,4}$,
B.~Jain$^{9}$,
E.~Krause$^{10,11}$,
T.~F.~Eifler$^{10,11}$,
D.~Gruen$^{4,5\ddag}$,
J.~Zuntz$^{12}$,
M.~A.~Troxel$^{1,2}$,
J.~Elvin-Poole$^{13}$,
J.~Prat$^{14}$,
M.~Wang$^{15}$,
S.~Dodelson$^{16}$,
A.~Kravtsov$^{17}$,
P.~Fosalba$^{7,8}$,
M.~T.~Busha$^{4}$,
A.~E.~Evrard$^{18,19}$,
D.~Huterer$^{19}$,
T.~M.~C.~Abbott$^{20}$,
F.~B.~Abdalla$^{21,22}$,
S.~Allam$^{15}$,
J.~Annis$^{15}$,
S.~Avila$^{23}$,
G.~M.~Bernstein$^{9}$,
D.~Brooks$^{21}$,
E.~Buckley-Geer$^{15}$,
D.~L.~Burke$^{4,5}$,
A.~Carnero~Rosell$^{24,25}$,
M.~Carrasco~Kind$^{26,27}$,
J.~Carretero$^{14}$,
F.~J.~Castander$^{7,8}$,
R.~Cawthon$^{17}$,
C.~E.~Cunha$^{4}$,
C.~B.~D'Andrea$^{9}$,
L.~N.~da Costa$^{24,25}$,
C.~Davis$^{4}$,
J.~De~Vicente$^{28}$,
H.~T.~Diehl$^{15}$,
P.~Doel$^{21}$,
J.~Frieman$^{15,17}$,
J.~Garc\'ia-Bellido$^{29}$,
D.~W.~Gerdes$^{18,19}$,
R.~A.~Gruendl$^{26,27}$,
G.~Gutierrez$^{15}$,
W.~G.~Hartley$^{21,30}$,
D.~Hollowood$^{31}$,
K.~Honscheid$^{1,2}$,
B.~Hoyle$^{32,33}$,
D.~J.~James$^{34}$,
T.~Jeltema$^{31}$,
D.~Kirk$^{21}$,
K.~Kuehn$^{35}$,
N.~Kuropatkin$^{15}$,
M.~Lima$^{36,24}$,
M.~A.~G.~Maia$^{24,25}$,
J.~L.~Marshall$^{37}$,
F.~Menanteau$^{26,27}$,
R.~Miquel$^{38,14}$,
A.~A.~Plazas$^{11}$,
A.~Roodman$^{4,5}$,
E.~Sanchez$^{28}$,
V.~Scarpine$^{15}$,
M.~Schubnell$^{19}$,
I.~Sevilla-Noarbe$^{28}$,
M.~Smith$^{39}$,
R.~C.~Smith$^{20}$,
M.~Soares-Santos$^{40}$,
F.~Sobreira$^{41,24}$,
E.~Suchyta$^{42}$,
M.~E.~C.~Swanson$^{27}$,
G.~Tarle$^{19}$,
D.~Thomas$^{23}$,
A.~R.~Walker$^{20}$,
J.~Weller$^{43,32,33}$
\begin{center} (DES Collaboration) \end{center}
}
\vspace{0.2cm}
}

\date{Accepted XXX. Received YYY; in original form ZZZ}

\pubyear{2018}

\begin{document}
\label{firstpage}
\pagerange{\pageref{firstpage}--\pageref{lastpage}}
\maketitle

\begin{abstract}
We use mock galaxy survey simulations designed to resemble the Dark Energy Survey Year 1 (DES Y1) data to validate and inform cosmological parameter estimation. When similar analysis tools are applied to both simulations and real survey data, they provide powerful validation tests of the DES Y1 cosmological analyses presented in companion papers. We use two suites of galaxy simulations produced using different methods, which therefore provide independent tests of our cosmological parameter inference. The cosmological analysis we aim to validate is presented in \citet{keypaper} and uses angular two-point correlation functions of galaxy number counts and weak lensing shear, as well as their cross-correlation, in multiple redshift bins.
While our constraints depend on the specific set of simulated realisations available, for both suites of simulations we find that the input cosmology is consistent with the combined constraints from multiple simulated DES Y1 realizations in the $\om-\sig$ plane. For one of the suites, we are able to show with high confidence that any biases in the inferred $S_8=\sig(\om/0.3)^{0.5}$ and $\om$ are smaller than the DES Y1 $1-\sigma$ uncertainties. For the other suite, for which we have fewer realizations, we are unable to be this conclusive; we infer a roughly 60\% (70\%) probability that systematic bias in the recovered $\om$ ($S_8$) is sub-dominant to the DES Y1 uncertainty. As cosmological analyses of this kind become increasingly more precise, validation of parameter inference using survey simulations will be essential to demonstrate robustness.
\end{abstract}
\blfootnote{$^{\star}$ E-mail: maccrann.2@osu.edu}
\blfootnote{$^{\dag}$ E-mail: jderose@stanford.edu}
\blfootnote{$^{\ddag}$ Einstein fellow}
\blfootnote{ Affiliations are listed at the end of the paper.}
\setcounter{footnote}{1}
\vspace*{4cm}





\section{Introduction}


The combination of our best cosmological datasets and our best theory of gravity supports our bizarre standard cosmological model: a Universe dominated by dark energy and dark matter. Dark energy is required to produce the observed acceleration of the Universe's expansion, and current observational constraints are consistent with the description of dark energy as a cosmological constant, $\Lambda$. In general, cosmological probes are sensitive to the properties of dark energy either because of its effect on the Universe's background properties (e.g. expansion rate, or average matter density as a function of cosmic time), or its effect on the growth of structure (or both). While some early indications that our Universe is not matter dominated came from galaxy clustering measurements sensitive to the latter (e.g. \citealt{maddox96}), arguably the most robust evidence for \lcdm\ ($\Lambda$ $+$ cold dark matter) comes from cosmological probes which are primarily sensitive to the former, known as \emph{geometrical probes}. The most mature of these are the distance-redshift relation of Type 1a supernovae (SN1a, e.g. \citealt{betoule14}), and the baryon acoustic peaks in the cosmic microwave background (e.g. \citealt{planckcosmo15}) and galaxy distributions (e.g \citealt{ross17,alam17}). Indeed the SN1a analyses of \citet{perlmutter99} and \citet{riess98} are considered the first convincing evidence of the Universe's late-time acceleration.

The development of more powerful probes of the growth of structure, as well as providing tighter constraints on the vanilla \lcdm\ model, are likely to be extremely useful for constraining deviations from \lcdm\ (e.g. \citealt{albrecht06,weinberg13}), for example models with time-evolving dark energy  and modified gravity, especially when combined with geometrical probes. Several observational programs are underway (The Dark Energy Survey\footnote{\url{https://www.darkenergysurvey.org}} (DES), The Kilo-Degree Survey\footnote{\url{http://kids.strw.leidenuniv.nl}} (KiDS), The Hyper Suprime-Cam Subaru Strategic Survey\footnote{\url{http://www.subarutelescope.org/Projects/HSC}}), that are designed to provide imaging in the optical and near infra-red that is sufficiently deep, wide and high quality to enable competitive cosmological information to be extracted from the Universe's large scale structure at $z<2$.
Meanwhile future surveys carried out by the Large Synoptic Survey Telescope\footnote{\url{http://www.lsst.org}} (LSST), Euclid\footnote{\url{http://sci.esa.int/science-e/www/area/index.cfm?fareaid=102}} and the Wide-Field Infrared Survey Telescope\footnote{\url{http://wfirst.gsfc.nasa.gov}} will enable order-of-magnitude improvements in cosmological constraints if systematic uncertainties can be controlled sufficiently.

Much of the information on structure growth available to these surveys lies well beyond the linear regime, so making theoretical predictions to capitalize on this information is challenging, because of computational expense (large N-body simulations are required), and because there exists theoretical uncertainty in how to implement the baryonic physics that affects the matter distribution on small scales \citep{schaye15,vogelsberger14}. Further modeling challenges arise when objects such as galaxies or clusters are used as tracers of the underlying matter field, since this requires understanding the statistical connection between these objects and the matter field. Cosmological simulations are crucial for tackling both of these challenges, and are already widely used to predict the clustering of matter on nonlinear scales \citep[e.g][]{smith03,heitmann10}. Some recent works have used cosmological simulations to directly make predictions for galaxy clustering statistics (e.g. \citealt{kwan15, sinha17}). 

The complexity of the analyses required to extract unbiased cosmological information from current and upcoming large-scale structure surveys demands thorough validation of the inference of cosmological parameters. Inevitably, approximations will be made in the model, for example to allow for fast likelihood evaluation in MCMC chains. While the impact of many of these can be investigated analytically (for example the impact of making the Limber approximation, or ignoring the effect of lensing magnification on galaxy clustering statistics), this requires the investigators to identify and  characterize all of these effects (and possibly their interactions) with sufficient accuracy. It would be complacent to ignore the possibility that some of these effects may slip through the net.

The modeling challenges described will be entangled with challenges related to the quality of the observational data such as spatially correlated photometric and weak lensing shear estimation errors, and photometric redshift uncertainties. It has been recognized that analysis of realistic survey simulations, which can naturally contain many of the theoretical and observational complexities of real survey data, will play a crucial part in this validation, for example both the Dark Energy Spectroscopic Instrument and LSST Dark Energy Science collaborations plan to complete a series of simulated data challenges before analysis of real survey data (e.g. \citealt{desc12}).
This is an especially powerful approach when one considers the importance of performing a blind analysis of the real survey data --- ideally one can finalize all analysis choices, informed by analysis of the survey simulations, before the analysis of the real data.

In this spirit, we use mock survey simulations for this task by attempting to recover the input cosmological parameters of the simulations using a methodology closely resembling that used on the real DES Year 1 (Y1) data in \citet{keypaper}. We note that since we are not directly using these simulations to provide theoretical predictions for the analysis of the real data, having simulations which match the properties of the real data in every aspect is not essential, although of course the more realistic the simulations are, the more valuable the validation they provide is. The simulations used in this work reflect the current state of the survey simulations used in the Dark Energy Survey, and are being improved as the survey progresses; we discuss some potential improvements in \sect{sec:discussion}. One of the challenges of such an analysis is disentangling biases in the inferred cosmological parameters caused by flaws in the inference process from those caused by features of the simulations that may not reflect the actual Universe. In this work we limit the amount of validation of the simulations themselves, in an effort to produce results on a similar timescale as the analyses of real DES Y1 data. 

This work considers two of the observables provided by galaxy imaging surveys: the weak gravitational lensing shear, and the galaxy number density. 
Probes of the growth of structure can be thought of as those which depend on the clustering statistics of the Universe's matter field; both the galaxy number density and the shear meet this requirement.
Weak gravitational lensing is the observed distortion of light emitted from distant sources by variations in the gravitational potential due to intervening structures. In galaxy imaging data, this manifests as distortions in the observed size, brightness and ellipticity of distant galaxies, which are referred to as \emph{source} galaxies. The ellipticity distortion is known as the \emph{shear}, and is the most commonly used weak lensing observable in galaxy surveys. 

Since the shear field depends on the projected matter density field (as well as the redshift of the source galaxies and the distance-redshift relation), its $N$-point statistics are directly sensitive to the $N$-point statistics of the intervening density field and the cosmological parameters that determine these. Cosmic shear alone can therefore provide competitive cosmological constraints \citep{bartelmann01,kilbinger15,shearcorr}.
Here we consider two-point shear correlations, which are primarily sensitive to the two-point correlation function of the matter over-density $\xi_{mm}(r)$. 

Galaxies meanwhile are assumed to reside in massive gravitationally bound clumps of matter often modeled as halos (spherical or ellipsoidal overdensities in the matter field). 
Thus, while galaxies trace the matter field (i.e. they are generally more likely to be found where there is more mass), they do so in a biased way: the overdensity (the fractional excess with respect to the mean) in number of galaxies at $\vect{x}$ is not the same as the overdensity in matter at $\vect{x}$. 
However, on sufficiently large scales we can assume linear biasing, such that the two-point correlation function of galaxies, 
$\xi_{gg}(r)$, can be related to the matter two-point correlation function via (e.g. \citealt{fry93})
\be
\xi_{gg}(r) = b_1^2 \xi_{mm}(r). \label{eq:lin_bias_2}
\ee 
The constant of proportionality, $b_1$, is known as the \emph{galaxy bias}, and depends on details of galaxy formation that most cosmological analyses do not attempt to model from first principles, instead leaving bias as a free nuisance parameter. In this case, galaxy clustering measurements alone (at least in the linear bias regime), do not provide strong constraints on the cosmologically sensitive matter clustering amplitude --- some other information is required to break the degeneracy with the galaxy bias.


The cross-correlation between galaxy number density and shear, also known as as \emph{galaxy--galaxy lensing}, can provide this information. It depends on the galaxy--matter cross-correlation, which in the  linear bias regime can also be related to $\xi_{mm}$ via
\be
\xi_{gm}(r) = b_1\xi_{mm}(r).
\ee
Hence galaxy clustering and galaxy-shear cross-correlations depend on complementary combinations of the galaxy bias and $\xi_{mm}$, and can be combined to allow useful cosmological inference (e.g. \citealt{mandelbaum13,kwan17}). 

This work is a companion to a cosmological parameter estimation analysis of Dark Energy Survey Year 1 (DES Y1) data, in which we use the three aforementioned two-point signals: cosmic shear, galaxy clustering and galaxy-galaxy lensing to infer cosmological parameters and test cosmological models \citep{keypaper}. Further details on the cosmic shear, galaxy clustering and galaxy-galaxy lensing parts of the analysis are available in \citet{shearcorr}, \citet{wthetapaper} and \citet{gglpaper} respectively. These are therefore the statistics we measure and model from the survey simulations considered in this work, in an attempt to demonstrate robust cosmological parameter inference.

In \sect{sec:theory} we describe the statistics estimated from the data, and how they are modeled. In \sect{sec:mocks}, we describe the suites of simulations used. In \sect{sec:samples}, we describe the galaxy samples used. In \sect{sec:results}, we present the correlation function measurements, our analysis choices (which closely follow those which are used on the real data in \citealt{keypaper}), and our inferred cosmological parameters. We also test the robustness of the constraints to photometric redshift errors.
We conclude with discussion in \sect{sec:discussion}.

\section{Two-point Statistics}\label{sec:theory}

We construct two galaxy samples --- one suited to estimating galaxy number density, and one suited to estimating shear. We will refer to the galaxy sample used to estimate number density as the \emph{lens} sample, and that used to estimate shear as the \emph{source} sample. From these two samples we construct three types of angular correlation functions - the auto-correlation of counts of the lens sample (galaxy clustering), the auto-correlation of the shear of the source sample (cosmic shear), and the cross-correlation between counts of the lens sample and shear of the source sample (galaxy--galaxy lensing).

The galaxy clustering statistic we use is $w(\theta)$, the excess number of galaxy pairs separated by angle $\theta$ over that expected from randomly distributed galaxies, estimated using the optimal and unbiased estimator of \citet{landy93}. 
Meanwhile, the information in galaxy--galaxy lensing is well-captured by the mean \emph{tangential shear}, $\left< \gamma_t(\theta) \right>$ ($\gamma_t(\theta)$ henceforth), the tangential component of the shear with respect to the lens-source separation vector, averaged over all lens-source pairs separated by angle $\theta$. In our estimation of the tangential shear, we include the subtraction of the tangential shear signal around points randomly sampled from the survey window function of the lens sample, which reduces the effects of additive shear biases (e.g. \citealt{hirata04}) and cosmic variance \citep{singh17}.

Since shear is a spin-2 field, one requires three two-point correlation functions to capture the two-point information of the shear field. One could use auto-correlations of the shear component tangential to the separation vector, $C_{++}(\theta)$, auto-correlations of the shear at \ang{45} to the separation vector, $C_{\times\times}(\theta)$, and the cross-correlation $C_{+\times}(\theta)$. In practice, $C_{+\times}(\theta)$ vanishes by parity arguments and we use the linear combinations of the remaining two correlations functions $\xipm (\theta) = C_{++}(\theta) \pm C_{\times\times}(\theta)$.

We split both lens and source galaxies into multiple bins in redshift, and measure correlation functions $\zeta^{ij}(\theta) \in \left\{ w^{ij}(\theta), \gamma_t^{ij}(\theta), \xip^{ij}(\theta), \xim^{ij}(\theta) \right\}$ between redshift bins $i$ and $j$. 

We use superscripts in the following to denote quantities relating to a particular redshift bin, so they should not be interpreted as exponents.
In general, an angular correlation function $\zeta_{\alpha\beta}^{ij}(\theta)$, can be related to a corresponding projected angular power spectrum, $C_{\alpha\beta}^{ij}(l)$ via 
\be
\zeta_{\alpha\beta}^{ij}(\theta) = \sum_{l} \frac{2l+1}{4\pi} C_{\alpha\beta}^{ij}(l) d^{l}_{mn}(\theta) \label{eq:cl2xi}
\ee
where $\alpha$ and $\beta$ represent the two quantities being correlated (galaxy overdensity $\delta_g$ or shear $\gamma$), and $d^{l}_{nm}(\theta)$ is the Wigner D-matrix. 
For the galaxy correlation function, $w(\theta)$, $m=n=0$, and the Wigner-D matrix reduces to the Legendre polynomial $P_l(\cos \theta)$. For the tangential shear, $\gamma_t(\theta)$, $m=2$ and $n=0$, and the Wigner-D matrix reduces to the associated Legendre polynomial $P^2_l(\cos \theta)$. For the shear correlation functions $\xipm (\theta)$, $m=2$ and $n=\pm2$; the Wigner D-matrices in this case can also be written in terms of associated Legendre polynomials (see \citealt{stebbins96} for the somewhat lengthy expressions).

In the small-angle limit, \eqn{eq:cl2xi} can be approximated with a Hankel transform
\be
\zeta_{\alpha\beta}^{ij}(\theta) = \int \dx{l} l C_{\alpha\beta}^{ij}(l) J_n(\theta), \label{eq:hankel}
\ee
where $n=0$ for $w(\theta)$, $n=2$ for $\gamma_t(\theta)$, $n=0$ for $\xip(\theta)$ and $n=4$ for $\xim(\theta)$. \citet{methodpaper} demonstrate that this approximation is sufficient for this analysis at the accuracy of DES Y1.

The angular power spectra, $C_{\alpha\beta}^{ij}(l)$ can be expressed in terms of the corresponding three-dimensional power spectra $P_{\alpha\beta}(k)$ as \citep{loverde08} 
\begin{align}
C_{\alpha\beta}^{ij}(l) =& \int_0^{\chi_h} d\chi D_A^{-1}(\chi) f_{\alpha}^i(\chi) J_{l+1/2}(k \chi) \nonumber \\
& \int_0^{\chi_h} d\chi' D_A^{-1}(\chi') f_{\beta}^j(\chi') J_{l+1/2}(k \chi') \nonumber \\
& \int_0^{\infty} dk k  P_{\alpha\beta}(k,\chi,\chi'). \label{eq:clfull}
\end{align}
$\chi$ is the comoving radial distance, $D_A(\chi)$ is the comoving angular diameter distance, $\chi_h$ is the horizon distance and $f_{\alpha}^i(\chi)$ and $f_{\beta}^i(\chi')$ are the appropriate projection kernels for computing the projected shear or number counts in redshift bin $i$ from the shear or number counts in three dimensions.

Under the Limber approximation \citep{limber} \eqn{eq:clfull} is simplified to
\begin{align}
C_{\alpha\beta}^{ij}(l) =& \int_0^{\chi_h} d\chi \frac{f_{\alpha}^i(\chi) f_{\beta}^i(\chi)}{D_A^2(\chi)} P_{\alpha\beta}(k=(l+1/2)/\chi,\chi). \label{eq:cllimber}
\end{align}

Predictions for each of the two-point correlation functions we use can therefore be derived using \eqnb{eq:hankel}{eq:cllimber} (in the flat-sky and Limber approximations); once we specify the appropriate power spectrum, $P_{\alpha\beta}(k,\chi)$, and two radial kernels, $f_{\alpha}^i(\chi)$ and $f_{\beta}^i(\chi)$.
For galaxy number counts, the projection kernel, $f_{\delta_g}(\chi)$ is simply the comoving distance probability distribution of the galaxy sample (in this case the lens sample) $n^i_{\mathrm{lens}}(\chi)$, normalised so that $\int d\chi n^i_{\mathrm{lens}}(\chi) = 1$. For shear the projection kernel for redshift bin $i$ is
\be
f_{\gamma}^i(\chi) = \frac{3 H_0^2 \Omega_m D_A(\chi)}{2c^2 a(\chi)} \int_{0}^{\chi_h} d\chi' n^i_{\mathrm{src}}(\chi)\frac{D_A(\chi'-\chi)}{D_A(\chi')},
\ee
where $n^i_{\mathrm{src}}(\chi)$ is the comoving distance probability distribution of the source galaxies. 

It follows that for $w^{ij}(\theta)$, the radial kernels are $n^i_{\mathrm{lens}}(\chi)$ and $n^j_{\mathrm{lens}}(\chi)$ and the power spectrum is the galaxy power spectrum, $P_{gg}^{ij}(k,\chi)$. In our fiducial model, we assume linear bias, and thus relate this to the matter power spectrum, $P_{mm}(k,z)$, via 
\be
P_{gg}^{ij}(k,z) = b_1^i b_1^j P_{mm}(k,z), 
\ee
where $b_1^{i}$ is a free linear galaxy bias parameter for redshift bin $i$, assumed to be constant over the redshift range of each lens redshift bin. In principle, there is also a shot noise contribution to the galaxy power spectrum. However, we neglect this term since any constant contribution to the power spectrum appears only at zero lag in the real-space statistics we use here, and we do not use measurements at zero-lag.

For $\gamma_t^{ij}(\theta)$, the radial kernels are $n^i_{\mathrm{lens}}(\chi)$ and $f_{\gamma}^j(\chi)$, and the appropriate power spectrum is the galaxy--matter power spectrum $P_{gm}(k,\chi)$, which in the linear bias regime is given by 
\be
P_{gm}^{ij}(k,\chi)=b_1^i P_{mm}(k,\chi).
\ee

Finally, for $\xi_{\pm}^{ij}(\theta)$, the radial kernels are $f_{\gamma}^i(\chi)$ and $f_{\gamma}^j(\chi)$, and the appropriate power spectrum is simply the matter power spectrum, $P_{mm}(k,\chi)$.

\section{Survey Simulations}\label{sec:mocks}

We now describe the two suites of simulations used in this work, which we will refer to as the \emph{BCC} and \emph{MICE}. The latter is already well-documented in \citet{fosalba15a,carretero15,fosalba15b}, hence we only include a brief description in \sect{sec:mice}. It will be useful in the following to note a few details of the DES Y1 dataset that is being simulated. The Year 1 dataset is constructed from DECam \citep{decam} images taken between August 2013 and February 2014 (see e.g. \citealt{y1gold}). An area of \SI{1786}{\deg^2} was imaged in $grizY$, but the cosmology analyses \citep{keypaper,shearcorr,gglpaper,wthetapaper} used only the contiguous \SI{1321}{\deg^2} region known as ``SPT'' \citep{y1gold}. 

\subsection{BCC simulations}

We make use of a suite of 18 simulated DES Year 1 galaxy catalogs constructed from dark matter-only N-body lightcones and include galaxies with DES \textit{griz} magnitudes with photometric errors appropriate for the DES Y1 data, shapes, ellipticities sheared by the underlying dark matter density field, and photometric redshift estimates. The N-body simulations were generated assuming a flat \lcdm\ cosmology with $\om=0.286$, $\Omega_b=0.047$, $n_s=0.96$, $h=0.7$ and $\sigma_8=0.82$. A more detailed description of this suite of simulations will be presented in \citet{derose2018}. These mocks are part of the ongoing `blind cosmology challenge' effort within DES, and hence are referred to as the \emph{BCC} simulations. 

\subsubsection{N-body Simulations}
For the production of large-volume mock galaxy catalogs suitable to model the DES survey volume, we use three different N-body
simulations per each set of 6 DES Year 1 catalogs. Any cosmological simulation requires a compromise between volume and resolution; the use of three simulation boxes per lightcone is intended to balance the requirements on volume and resolution which change with redshift. At lower redshift, less volume is required for the same sky area compared to higher redshift, but higher resolution is required to resolve the excess nonlinear structure on a given comoving scale. 
Properties of the three
simulations are summarized in \tab{table:simulations}. 
All simulations are run using the code L-Gadget2,
a proprietary version of the Gadget-2 code  \citep{gadget2} optimized for memory efficiency and designed explicitly to run large-volume dark
matter-only N-body simulations.  

Additionally, we have modified this code to create a particle lightcone output on the fly. Linear power spectra computed with 
CAMB \citep{lewis:04} were used with 2LPTic \citep{crocce_etal:06} to produce the initial conditions using second-order Lagrangian perturbation theory. 
\begin{table*}
\caption{Description of the N-body simulations used.}
\begin{tabular}{ccccccccc}
Simulation & box size & particle number & mass resolution & force resolution & halo mass cut\\
BCC ($0.00<z<0.34$) & \SI{1.05}{\hinv\gigaparsec} & 1400$^3$ & \SI{2.7e10}{\hinv\msun} & \SI{20}{\hinv \kilaparsec} & \SI{3.0e12}{\hinv\msun}\\
BCC ($0.34<z<0.90$) & \SI{2.60}{\hinv\gigaparsec} &  2048$^3$& \SI{1.3e11}{\hinv\msun} & \SI{35}{\hinv \kilaparsec} & \SI{3.0e12}{\hinv\msun}\\
BCC ($0.90<z<2.35$) & \SI{4.00}{\hinv\gigaparsec} & 2048$^3$& \SI{4.8e11}{\hinv\msun} & \SI{53}{\hinv \kilaparsec} & \SI{2.4e13}{\hinv\msun}\\
MICE & \SI{3.07}{\hinv\gigaparsec} & 4096$^3$& \SI{2.93e10}{\hinv\msun} & \SI{50}{\hinv \kilaparsec} & $\sim \SI{e11}{\hinv\msun}$ \\
\label{table:simulations}
\end{tabular}
\end{table*}

\subsubsection{Galaxy Model}\label{sec:addgals}

Galaxy catalogs are built from the lightcone simulations using the \textsc{ADDGALS} algorithm. We briefly describe the algorithm, and refer the reader to \citet{addgals, derose2018} for more details. 

The main strengths of this algorithm are its ability to reproduce the magnitude-dependent clustering signal found in subhalo abundance matching  (SHAM) models, and its use of empirical models for galaxy SEDs to match color distributions. SHAM models have been shown to provide excellent fits to observed clustering data \citep{conroy06, lehmann17}, thus by matching SHAM predictions \textsc{ADDGALS} is able to accurately reproduce observed clustering measurements as well. 

The \textsc{ADDGALS} algorithm can be subdivided into two main parts. First, using a SHAM on a high-resolution $N$-body simulation, we fit two independent parts of the galaxy model: $p(\delta|M_{r},z)$, the distribution of matter overdensity, $\delta$, given galaxy absolute magnitude, $M_{r}$, and redshift, $z$, and $p(M_{r,\mathrm{cen}} | M_{\mathrm{halo}}, z)$, the distribution of $r$-band absolute magnitude of central galaxies, $M_{r,\mathrm{cen}}$, given host halo mass, $M_{\mathrm{halo}}$, and redshift. To do this we subhalo abundance match a luminosity function $\phi(M_{r}, z)$, which has been constrained to match DES Y1 observed galaxy counts, to 100 different redshift snapshots and measure $\delta$ centered on every galaxy in the SHAM. The model for $p(\delta|M_{r},z)$ is then fit to histograms of $\delta$ in narrow magnitude bins in the SHAM in each snapshot. The model for $p(M_{r,\mathrm{cen}} | M_{\mathrm{halo}}, z)$ is similarly constrained by fitting to the distributions of $M_{r,\mathrm{cen}}$ in bins of $M_{\mathrm{halo}}$ for each snapshot. \citet{addgals} shows that reproducing these distributions is sufficient to match the projected clustering found in the SHAM.

Now, using $\phi(M_{r},z)$, $p(\delta|M_{r},z)$, and $p(M_{r,\mathrm{cen}} | M_{\mathrm{halo}}, z)$, we add galaxies to our lightcone simulations. Working in redshift slices spanning $z_{\mathrm{low}}<z\le z_{\mathrm{high}}$, we first place galaxies on every resolved central halo in the redshift shell, where the mass of a resolved halo, $M_{min}$ ,is given in \tab{table:simulations}, drawing its luminosity from $p(M_{r,\mathrm{cen}} | M_{\mathrm{halo}}, z)$. As these simulations are relatively low resolution, this process only accounts for a few percent of the galaxies that DES observes. For the rest, we create a catalog of galaxies with absolute magnitudes $\{M_{r,i}\}$ and redshifts $\{z_{i}\}$, with $i=1,\ldots,N$ and $N=\int_{z_{\mathrm{low}}}^{z_{\mathrm{high}}}dz\frac{dV}{dz}\phi_{\mathrm{unres}}(M_{r},z)$, where 
\begin{align}
\phi_{\mathrm{unres}} (M_{r},z) =& \phi (M_{r},z) - \phi_{\mathrm{res}} (M_{r},z) \\
=& \phi (M_{r},z) \\ 
&-\int_{M_{min}}^{\infty}dM_{\mathrm{halo}}p(M_{r,cen}|M_{\mathrm{halo}},z)n(M_{\mathrm{halo}},z).
\end{align}
Each $M_{r,i}$ is drawn from $\phi_{\mathrm{unres}}(M_{r},z_{\mathrm{mean}})$, where $z_{\mathrm{mean}}$ is the mean redshift of the slice, and each $z_{i}$ is drawn uniformly between $z_{\mathrm{low}}<z_{i}\le z_{\mathrm{high}}$. It can be shown that this uniform distribution is appropriate, since the distribution of particles in the lightcone already accounts for the change in comoving volume element as a function of redshift, $\frac{dV}{dz}$. Finally, in order to determine where to place each galaxy, we draw densities $\{\delta_{i}\}$ from $p(\delta|M_{r,i}, z_{i})$, and assign the galaxies to particles in the lightcone with the appropriate density and redshift. Notably, we make no explicit classification of galaxies as centrals or satellites when they assigned in this way.

Once galaxies have been assigned positions and $r$-band absolute magnitudes, we measure the projected distance to their fifth nearest neighbor in redshift bins of width $\Delta z = 0.05$. We then bin galaxies in $M_{r}$ and rank order them in terms of this projected distance. We compile a training set consisting of the magnitude-limited spectroscopic SDSS DR6 VAGC cut to $z<0.2$ and local density measurements from \cite{cooper2006}. This training set is rank ordered the same way as the simulation. Rank ordering the densities allows us to use a non-volume limited sample in the data, since this rank is preserved under the assumption that galaxies of all luminosities are positively biased. Each simulated galaxy is assigned the SED from the galaxy in the training set with the closest density rank in the same absolute magnitude bin. The SED is represented as a sum of templates from \cite{blanton2003}, which can then be used to shift the SED to the correct reference frame and generate magnitudes in DES band passes. While the use of this training set neglects the evolution of the relationship between rank local density and SED between $z<0.2$ and the higher redshifts probed by DES, it does provide a sample with high completeness over the required range of galaxy luminosity. The use of rank density should reduce the amount of redshift evolution in this relationship, but residual effects may be present in the color dependent clustering of the BCC simulations. While the agreement between redMaGiC angular clustering in the data and the BCC simulations is generally good, there are some redshift dependent differences that could be partially attributable to this effect \citep{derose2018}. Planned improvements of the algorithm will take advantage of higher redshift spectroscopic datasets. 

Galaxy sizes and ellipticities are assigned based on the galaxies' observed $i$-band magnitude based on fits to the joint distribution of these quantities in high resolution Suprime-Cam data \citep{suprimecam}.

\subsubsection{Raytracing}
In order to derive weak lensing quantities for each galaxy, we employ a multiple-plane raytracing algorithm called {\sc Calclens} \citep{becker2013}. The raytracing is done on an $n_{side}=4096$ \emph{HEALPIX} \citep{healpix} grid, leading to an angular resolution of approximately $0.85'$. At each lens plane, the Poisson equation is solved using a spherical harmonic transform, thus properly accounting for sky curvature and boundary conditions. The inverse magnification matrix is interpolated from each ray at the center of each lens plane to the correct angle and comoving distance of each galaxy. The magnitudes, shapes and ellipticities of the galaxies are then lensed using this information. 

\subsubsection{Photometric Errors and Footprint}\label{sec:depthmap}
To create each of the BCC Y1 catalogs, a rotation is applied to the simulated galaxies to bring them into the DES Y1 SPT footprint described in \sect{sec:mocks}. The DES Y1 mask is applied and the area with $RA<0$ is cut in order to fit 6 Y1 footprints into each simulation, leaving an area of \SI{1122}{\deg^2} out of the original \SI{1321}{\deg^2}. Applying this cut allows us to use more area in each simulated half-sky (without the cut we are only able to fit 2 Y1 footprints into each simulated half-sky without overlap), and therefore allows us to test our cosmological parameter inference with greater statistical precision.
 Photometric errors are applied to the \emph{BCC} catalogs using the DES Y1 Multi Object Fitting (\emph{MOF}) depth maps. The errors depend only on the true observed flux of the galaxy and its position in the footprint, and not its surface brightness profile. 

\subsection{MICE simulations}\label{sec:mice}

We use the MICE Grand Challenge simulation (MICE-GC), which is well documented in \citet{fosalba15a,carretero15,fosalba15b}; we provide a brief description here for convenience.
MICE-GC constitutes a \SI{3}{\gigaparsec/h} N-body simulation with $4096^3$ particles, produced using the Gadget-2 code \citep{gadget2} as described in \citet{fosalba15b}. The cosmological model is flat \lcdm\ with $\om=0.25$, $\Omega_b=0.044$, $n_s=0.95$, $h=0.7$ and $\sigma_8=0.8$.  The mass resolution is \SI{2.93e10}{\hinv\msun} and the force softening length is \SI{50}{\hinv\kilaparsec}. Halos are identified using a Friends-of-Friends algorithm (with linking length 0.2 times the mean inter-particle distance) and these are populated with galaxies via a hybrid sub-halo abundance matching (SHAM) and halo occupation distribution (HOD) approach \citep{carretero15} designed to match the joint distributions of luminosity, $g-r$ color, and clustering amplitude observed in SDSS \citep{blanton2003,zehavi05}. Weak lensing quantities are generated on a \emph{HEALPIX} grid of $N_{\mathrm{side}}=8192$ (an angular resolution of $\approx0.4'$) assuming the Born approximation (see \citealt{fosalba15a} for details).

We rotate the MICE octant into the DES Y1 footprint and imprint the spatial depth variations in the real DES Y1 data onto the MICE galaxy magnitudes using the same method as for the BCC (see \sect{sec:depthmap}). We find we can apply two such rotations which retain the majority of the Y1 area and have little overlap in the Y1 area. Hence we have two MICE-Y1 realisations. 

\subsection{Notable differences between the mock catalogs}
We note the following significant differences between the mock catalogs constructed from the BCC and MICE simulations:
\begin{itemize}
\item{Volume of data: We have 18 DES Y1 realisations for the BCC simulations, in principle allowing a measurement of any bias in the recovered cosmological parameters with uncertainty $1/\sqrt{18}$ of the DES Y1 statistical error. We note that the slightly smaller area used for the BCC simlations will result in a small loss of constraining power.
For MICE on the other hand, we expect uncertainty on the recovered parameters that is more comparable to the DES Y1 statistical errors 
(a factor of $1/\sqrt{2}$ smaller).
Ideally, the uncertainty on the inferred parameter biases should be subdominant to the achieved parameter constraint for DES Y1. Clearly what constitutes `subdominant' is somewhat subjective, but we consider the 18 BCC realisations as satisfactory in this respect, while more MICE realisations would be desirable to satisfy this requirement.}
\item{Each BCC realisation is constructed from three independent simulation boxes, resulting in discontinuities in the density field where they are joined together, while MICE uses a single box.}
\item{Resolution: The mass resolution of the lowest redshift BCC simulation box ($\SI{2.7e10}{\h^{-1}\msun}$) is similar to MICE ($\SI{2.93e10}{\h^{-1}\msun}$) with significantly higher force resolution, while the higer redshift boxes have signifcantly lower mass resolution (see \tab{table:simulations}) and comparable force resolution.}
\item{Galaxies are added to the N-body simulations using different methods (BCC uses \textsc{ADDGALS}, while MICE uses a hybrid SHAM and HOD approach); in general this will lead to different galaxy bias behaviour in the nonlinear regime.}
\item{Weak lensing quantities in BCC are calculated using full ray-tracing, whereas in MICE they are calculated under the Born approximation. We do not however expect this difference to be significant for the relatively large-scale observables considered here, indeed we do not include beyond-Born approximation contributions in the theoretical modeling of the lensing signals used in our cosmological parameter inference.}
\item{For BCC, we use BPZ \citep{benitez00} photometric redshift estimates, the fiducial \photoz\ method used for the weak lensing source galaxies on the real DES Y1 data. For MICE, we use true redshifts for the weak lensing galaxies throughout.}
\end{itemize}

\section{Galaxy samples}\label{sec:samples}

\begin{figure*}
\includegraphics[width=\columnwidth]{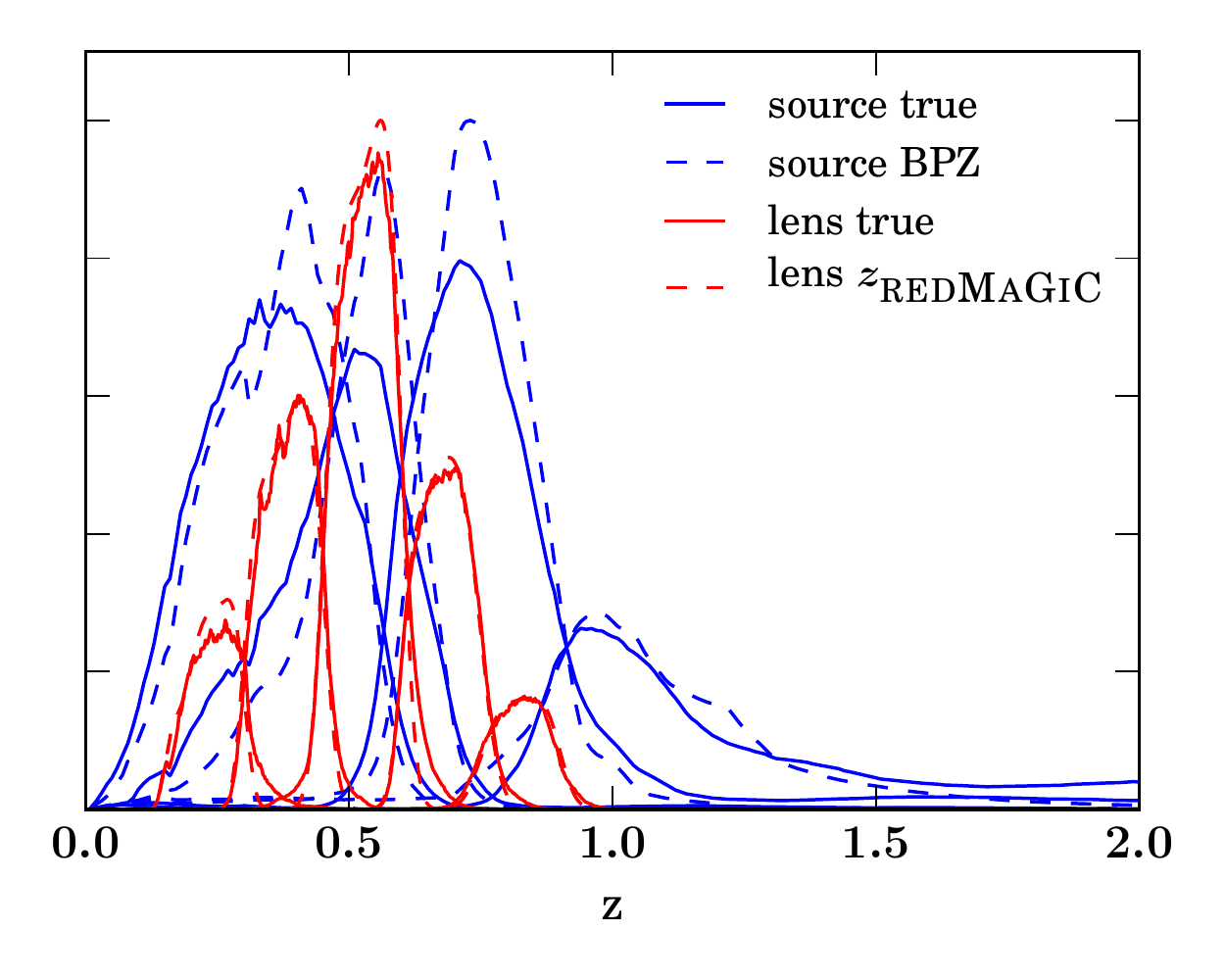}
\includegraphics[width=\columnwidth]{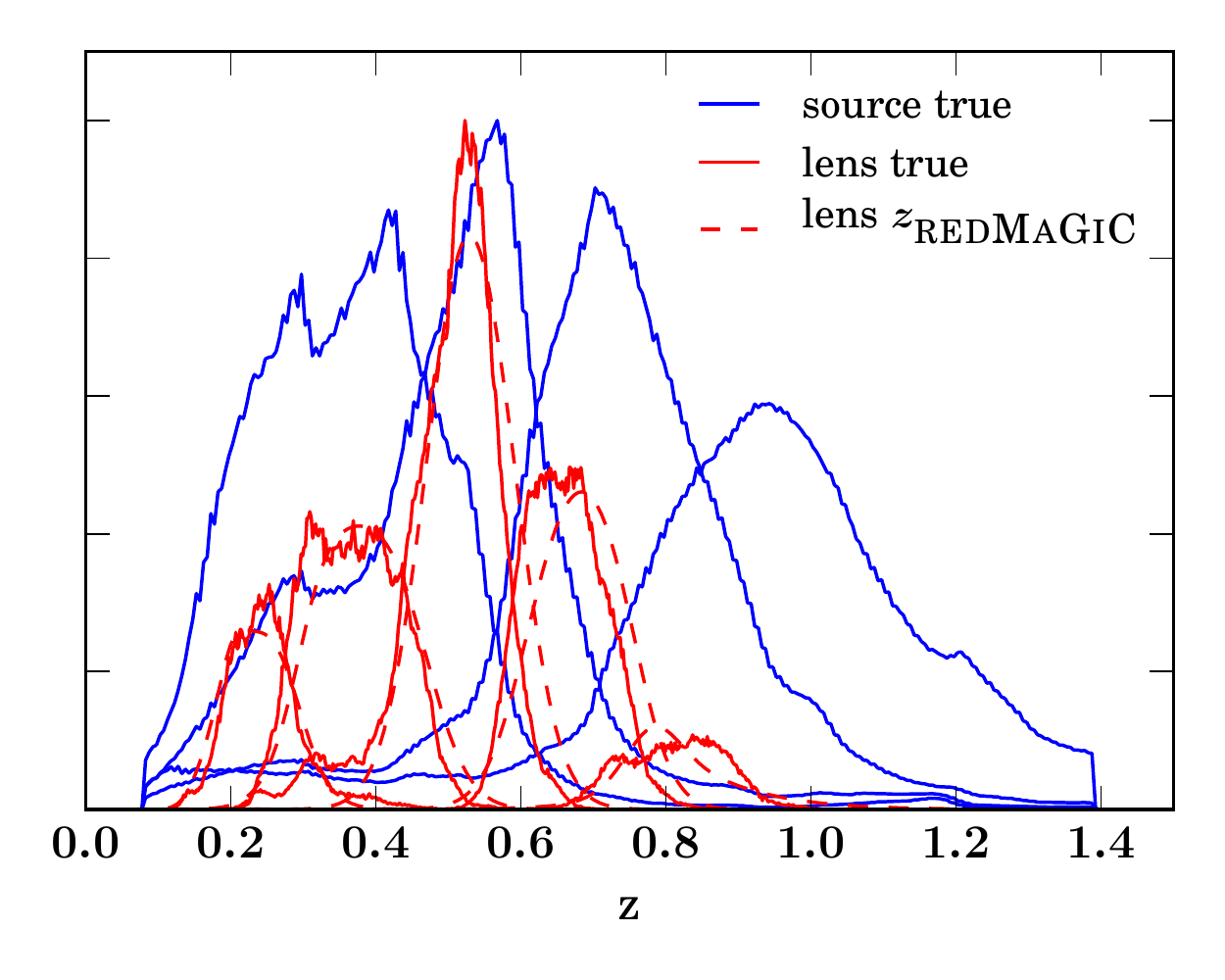}
\caption[]{Redshift distributions for the galaxy samples used. Red and blue indicate the \redmagic\ (\emph{lens}) galaxies and the weak lensing source galaxies respectively. Solid and dashed lines indicate the true distributions and those estimated from photometric redshifts respectively. Left panel: BCC, right panel: MICE. As discussed in \sect{sec:wlsample}, we do not have BPZ \photoz\ estimates for the weak lensing source sample in MICE and by construction the true source redshift distributions match the BPZ estimates for the real data, apart from above $z=1.4$, the maximum redshift of the MICE galaxies. For visual clarity, the lens and source redshift distributions have arbitrary normalization.
}
\label{fig:nzs}
\end{figure*}

We select two different galaxy samples from the mock catalogs, chosen based on their suitability to probe the galaxy number density (and act as the lens sample for the galaxy--galaxy lensing) and weak lensing shear fields respectively. 

\subsection{Lens sample}
To probe the galaxy number density, we use galaxies selected using the \redmagic\ algorithm \citep{redmagicSV}. \redmagic\ fits an empirically calibrated red-sequence template to all objects, and then selects those which exceed some luminosity threshold (assuming the photometric redshift inferred from the red-sequence template fit), and whose colors provide a good fit to the red-sequence template. This allows selection of a bright, red, galaxy sample with approximately constant comoving number density. The fact that they are close to the red sequence allows high quality photometric redshift (\photoz) estimation --- the \redmagic\ galaxies used in the DES Y1 analyses \citep{wthetapaper,keypaper} have an average standard error, $\sigma_z = 0.017(1+z)$. We refer to the \photoz s estimated by the \redmagic\ algorithm as $\zred$. 

As in \citet{wthetapaper} and \citet{keypaper}, we split the \redmagic\ sample into five redshift bins, defined
$0.15<\zred<0.3$, 
$0.3<\zred<0.45$, 
$0.45<\zred<0.6$, 
$0.6<\zred<0.75$, 
$0.75<\zred<0.9$.
For the first three redshift bins, the \redmagic\ \emph{high density} sample is used (luminosity, $L>0.5L_*$; number density, $n_{\mathrm{gal}} = \SI{4e-3}{\megaparsec^{-3}}$), while the fourth and fifth redshift bins are selected from the high luminosity ($L>L_*$, $n_{\mathrm{gal}} = \SI{1e-3}{\megaparsec^{-3}}$) and higher luminosity ($L>1.5L_*$, $n_{\mathrm{gal}} = \SI{1e-4}{\megaparsec^{-3}}$) samples respectively . Selecting brighter galaxies at high redshift allows the construction of a sample with close to uniform completeness over a the majority of the DES Year 1 footprint.
The true redshift distributions\footnote{The comoving galaxy number density as a function of redshift.} ($n(z)$s henceforth) of the \redmagic\ galaxies are shown as the red solid lines in \fig{fig:nzs}. These are histograms of the true redshift ($\ztrue$) for all galaxies within a given bin. 

The $n(z)$s can also be estimated using, \zred, and the associated uncertainty, $\sigma(\zred)$, which is also provided by the \redmagic\ algorithm. These quantities are designed such that the probability of a \redmagic\ galaxy having true redshift $\ztrue$, $p(\ztrue \vert \zred)$ is well approximated by a Gaussian distribution with mean and $\sigma$ given by \zred\ and $\sigma(\zred)$ respectively. Thus the redshift distribution of each \redmagic\ tomographic bin can be estimated  by stacking this Gaussian $p(\ztrue \vert \zred)$ estimate over all objects in that bin. This is how the \redmagic\ $n(z)$s are estimated on the real data, where true redshifts are not available, and is shown as the dashed red lines in \fig{fig:nzs}.

For BCC, the visual agreement is good although there are some differences, for example the $n(z)$ looks to be underestimated at the high redshift end of the first three redshift bins. For MICE, the \redmagic\ \photoz\ estimate also performs well, although some bias is apparent for the highest three redshift bins. Averaged over all simulations, there are 580000 galaxies in our lens sample in BCC, compared with 660000 in the DES Y1 data. Accounting for the difference in the areas of the footprints, these numbers agree to $5\%$ accuracy. This result also holds for the MICE lens sample, which contains 590000 galaxies.


\subsection{Weak lensing source sample}\label{sec:wlsample}

Unlike for galaxy clustering measurements, for shear correlation function measurements, a galaxy sample whose completeness varies across the sky can be used, since number density fluctuations are not the quantity of interest. Instead we require the sample to provide an unbiased estimate of the shear in any region of the sky. As in \sect{sec:theory}, we call this sample the source sample.
We note that fluctuations in the galaxy number density can produce higher order effects on weak lensing statistics (see e.g. \citealt{hamana02,schmidt09}), but are below the few percent level for the angular scales used here \citep{maccrann17}. For the real DES Y1 data, the weak lensing source selection depends on the outputs of 5+ parameter model-fitting shear estimation codes. To simulate this selection would require propagating our mock galaxy catalogs into image simulations with realistic galaxy appearances, which is beyond the scope of this work. 

For the BCC, we perform cuts on galaxies' signal-to-noise and size relative to the point-spread-function (PSF) (assuming the spatial noise and PSF size distributions observed in the DES Y1 data) that yield a sample with similar number density as the weak lensing sample in the real data. 
Specifically, we make the following cuts:
\begin{enumerate}
\item Mask all regions of the footprint where limiting magnitudes and PSF sizes cannot be estimated.
\item $m_r < -2.5 \log_{10}(1.5) + m_{r, lim}$
\item $\sqrt{r_{gal}^{2} + (0.13 r_{PSF})^{2}} > 1.25r_{PSF}$
\item $m_r < 22.01 + 1.217 z$
\end{enumerate}
where $m_{r,lim}$ and $r_{PSF}$ are the limiting magnitude and PSF FWHM estimated from the data at the position of each galaxy. The first two cuts approximate signal to noise related cuts that are be applied to shape catalogs in the data. Using only these, the BCC simulations yield number densities that exceed those found in the data, so also apply the third cut in order to more closely match the DES Y1 shape noise.


We then use the provided BPZ \citep{benitez00} photometric redshift estimates (BPZ is the fiducial method used to estimate photometric redshifts of the source galaxies in the real DES Y1 data, see \citealt{photoz}) to split the source sample into redshift bins. As in \citet{keypaper} we split the weak lensing sample 
into four redshift bins, based on the mean of the BPZ redshift PDF, $\zm$. Given that the size of the photometric redshift uncertainties is comparable to the bin widths, there is little to be gained by using more redshift bins. The four redshift bins are defined 
$0.2<\zm<0.43$, $0.43<\zm<0.63$, $0.63<\zm<0.9$,
$0.9<\zm<1.3$.
The $n(z)$s of the source sample are shown as the blue lines in \fig{fig:nzs}. The histograms of true redshift for each bin are shown as solid lines, and $n(z)$s estimated using the BPZ redshift PDF estimates are shown as dashed lines. Again some mis-estimation of the true $n(z)$s is apparent; we assess the impact of this in \sect{sec:photozs}. 

We do not have photometric redshift estimates for the MICE catalogs, so instead randomly sample MICE galaxies to produce the same tomographic $n(z)$s as estimated by BPZ on the real data. In detail, we take the BPZ $n(z)$ estimates for the source sample from \citet{photoz}, and for objects at a given true redshift in the MICE catalogs, we randomly assign a redshift bin with probability given by the relative amplitude of each tomographic $n(z)$ at that redshift. We additionally assign the MICE galaxies weights so that the weighted $n(z)$ for each tomographic bin matches the shape of the BPZ $n(z)$ (within the redshift range of the MICE galaxies, which does not extend above $z=1.4$). The resulting $n(z)$s are shown in the right panel of \fig{fig:nzs}.

We add Gaussian-distributed shape noise to the MICE source sample galaxies such that $\sigma_e^2/n_{\mathrm{eff}}$, (where $\sigma_e$ is the ellipticity dispersion, and $n_{\mathrm{eff}}$ is the effective galaxy number per unit area) matches the DES Y1 data. This ensures the covariance of the lensing statistics have the same shape noise contribution as the real DES Y1 data. Averaged over all BCC simulations, there are 23 million galaxies, compared to 26 million in the Y1 data. Taking into account the differences in area, these agree to $5.5\%$ accuracy. 

\section{Results}\label{sec:results}

In this section we present measurements of the two-point correlation functions described in \sect{sec:theory} on the galaxy samples described in \sect{sec:samples}. We then summarise the choices made for the analysis of these measurements, and finally present cosmological parameter constraints, and discuss how these should be interpreted.

\subsection{Measurements and covariance}\label{sec:meas}

We estimate the two-point correlation functions using \textsc{treecorr}\footnote{\url{https://github.com/rmjarvis/TreeCorr}} \citep{jarvis04}. We compute correlation functions for all redshift bin combinations i.e. fifteen combinations for $\wtheta$, 20 combinations for $\gammat$, and 10 combinations for $\xipm(\theta)$. We compute the correlation functions in 20 log-spaced angular bins in the angular range $2.5 < \theta < 250$ arcminutes.

We show in \app{app:tpt_meas} all the two-point correlation function measurements used. Figures~\ref{fig:xip}-\ref{fig:wtheta} show the two-point measurements on the BCC sims. 
Figures~\ref{fig:micexip}-\ref{fig:micewtheta} are the corresponding plots for the MICE-Y1 catalogs. Shaded regions indicated angular scales not used in the fiducial cosmological analysis because of theoretical uncertainties in the non-linear regime.

For all individual two-point functions and their combinations, we use the covariance matrix presented in \citet{methodpaper}, which uses an analytic treatment of the non-Gaussian terms \citep{eifler14,krause16} based on the halo model \citep{seljak00,peacock00}. We calculate the covariance assuming the true cosmology for each simulation. This is clearly not possible in an analysis of real data, where using an incorrect assumed cosmology (or in fact, not including the parameter dependence of the covariance matrix) could potentially introduce parameter biases. However, \citet{keypaper} did demonstrate there was negligible change in the parameter constraints when using two different cosmologies to calculate the covariance matrix, so we do not believe our conclusions are very sensitive to this choice. 

In the covariance calculation, we do not include the survey geometry corrections to the pure shape or shot noise covariance terms discussed in \citet{troxel18}. For the DES Y1 geometry, the correction to the pure shot and shape noise contributions to the covariance are at most $\sim20\%$, and this is at the largest scales, where shot/shape noise is generally subdominant.


We also do not include redshift bin cross correlations in $w(\theta)$, since we do not expect the fiducial theoretical model used, which assumes the Limber approximation and does not include redshift space-distortions or magnification contributions, to be sufficiently accurate for these parts of the data vector (see e.g. \citealt{loverde08}, \citealt{montari15} for the importance of not using the Limber approximation, and including magnification respectively for widely separated redshift bins). 

\subsection{Analysis choices}\label{sec:analysischoices}


We summarize below our analysis choices, which closely follow those of \citet{methodpaper} and \citet{keypaper}, where the methodology and the application to data of the DES Y1 key cosmological analysis are described.

\begin{enumerate}
\item{\textbf{Gaussian Likelihood}. We assume the measured datavectors are multivariate-Gaussian distributed, with the covariance matrix described in \sect{sec:meas}. We note this is an approximation (see e.g. \citealt{sellentin18}); but any impact on parameter constraints will be mitigated by the significant contributions of shot noise and shape noise to the covariance matrix.}
\item{\textbf{Minimum angular scales.} For \wtheta\ and \gammat\ we use minimum angular scales corresponding to \SI{8}{\hinv\megaparsec} and \SI{12}{\hinv\megaparsec} at the mean redshift of the lens redshift bin, respectively (following \citealt{keypaper,methodpaper}). These minimum scales are justified in \citet{methodpaper}, who studied the potential impact of ignoring nonlinear galaxy bias on the inferred cosmological parameters. This was estimated by generating fake DES Y1-like datavectors which included analytic models for nonlinear galaxy bias, which were used as input to a cosmological parameter estimation pipeline that assumed linear bias. The minimum scales were chosen such that biases in cosmological parameters were small compared to the uncertainties on those parameters. 
The analysis of galaxy simulations in this work provides a further test of the effectiveness of these scale cuts. 

For $\xipm$ we use the same minimum angular scales as \citet{shearcorr} and \citet{keypaper}, where we use the following  procedure: for each redshift bin combination we calculate the fractional difference in the expected signal when the matter power spectrum prediction used is modulated using templates from the OWLS simualations \citep{schaye10}. Separately for \xip\ and \xim\, and for each redshift bin combination, we cut all angular scales smaller than and including the largest angular scale where the fractional difference exceeds 2\%. While this scale cut was motivated in \citet{shearcorr} by the possibility of systematic biases due to baryonic physics not included in the simulations used here, we use it since removing these small scales will reduce the impact of finite simulation resolution on the cosmic shear signal.}
\item{\textbf{Galaxy bias model} As in the real data analysis \citep{keypaper}, we marginalize over a single linear bias parameter, $b^i_1$ per lens redshift bin $i$. We assume no redshift evolution of the bias across each redshift bin, but have verified that assuming passive evolution within a redshift bin i.e.\ $b^i(z) \propto D(z)$, where $D(z)$ is the linear growth factor, produces negligible differences in our parameter constraints.
}
\item{\textbf{Redshifts.} For the results in \sect{sec:cosmo} we use true redshifts to construct the $n(z)$s for the theory predictions. As discussed in \sect{sec:photozs}, we find indications that the performance of BPZ on the BCC simulations is significantly worse than on the real data. Therefore, while we still use BPZ point redshift estimates to place galaxies in tomographic bins throughout, in \sect{sec:cosmo} we show constraints which use true redshift information to construct the $n(z)$ (which enters the projection kernels $f_{\alpha}(\chi)$ in \eqn{eq:cllimber}).}
\item{\textbf{Matter power spectrum.} Following \citet{keypaper}, we use \camb\ to calculate the linear matter power spectrum and  \halofit\ \citep{smith03,takahashi2012} to model the nonlinear matter power spectrum.}
\item{\textbf{Limber approximation.} We use the Limber approximation to calculate all angular power spectra and do not include the contributions from redshift-space distortions; \citet{methodpaper} demonstrate that this is sufficiently accurate for DES Y1.}
\item{\textbf{Free parameters.} As well as five linear galaxy bias parameters, $b_1^i$, we marginalize over the same set of cosmological parameters (and use the same priors) as in \citet{keypaper}, with the exception of the sum of neutrino masses, $\sum m_{\nu}$. Since $\sum m_{\nu}=0$ in both simulations suites, using a prior of $\sum m_{\nu}>0$, would inevitably bias the inferred $\sum m_{\nu}=0$ high, and given it is degenerate with other cosmological parameters, this would bias the inference of the other cosmological parameters. We also do not include nuisance parameters designed to account for effects not present in the simulations (so unlike the \citealt{keypaper}, we do not maginalise over intrinsic alignment parameters or shear calibration uncertainties).}
\end{enumerate}

\subsection{Fiducial cosmological parameter constraints}\label{sec:cosmo} 

Having made measurements from all simulation realizations, and defined a modeling framework to apply to them, it is worth taking a step back to think about what information we wish to extract. Our aim is to estimate systematic biases in inferred parameters due to failures in our analysis and modeling of the simulations. We note that of course we will only be sensitive to those sources of systematic biases that are present in the simulations. For example, neither simulation suite here includes galaxy intrinsic alignments (and we do not include this effect in our modeling). Furthermore, as noted in \sect{sec:analysischoices}, we remove the effect of photometric redshift biases for the results shown in this section, and use true redshift information (we discuss the photometric redshift performance for the BCC simulations in \sect{sec:photozs}).

We estimate the size of systematic biases in our inferred parameters in the following way. We assume $P^{\mathrm{sys}}(\vect{\theta}, s_i)$, the potentially systematically biased posterior on parameters $\vect{\theta}$ we infer from a simulated datavector $s_i$ is related to the true posterior by some constant translation in parameter space:
\be
P^{\mathrm{sys}}(\vect{\theta}, s_i) = P(\vect{\theta} - \Delta\vect{\theta} | s_i). \label{eq:psys}
\ee
We wish to estimate the posterior on $\Delta\vect{\theta}$. We start by considering $P(s_i | \vect{\theta}, \Delta\vect{\theta} )$, the probability of drawing simulated datavector $s_i$ given a value of $\Delta\vect{\theta}$, and a set of true parameters (i.e. those input to the simulation), $\vect{\theta}_{\mathrm{true}}$. This probability is independent of $\Delta\vect{\theta}$ such that 
\begin{align}
P( s_i |  \vect{\theta}_{\mathrm{true}}, \Delta\vect{\theta} ) &= P( s_i | \vect{\theta}_{\mathrm{true}}) \\
& = \frac{ P( \vect{\theta}_{\mathrm{true}} | s_i ) P(s_i) }{ P( \vect{\theta}_{\mathrm{true}} ) } \\
& = \frac{ P^{\mathrm{sys}}( \vect{\theta}_{\mathrm{true}} + \Delta\vect{\theta}, s_i ) P(s_i) }{ P( \vect{\theta}_{\mathrm{true}} ) } \label{eq:psi}
\end{align}
where in the second line we have used Bayes' theorem, and we have substituted \eqn{eq:psys} in the third line. We can again use Bayes' theorem to rewrite the left-hand side:
\begin{align}
P( \Delta\vect{\theta} | s_i, \vect{\theta}_{\mathrm{true}} ) = \frac{P( s_i |  \vect{\theta}_{\mathrm{true}}, \Delta\vect{\theta} ) P(\Delta\vect{\theta}) }{ P(s_i) }.
\end{align}
Substituting \eqn{eq:psi}, and assuming a flat prior $P(\Delta\vect{\theta})$, we have
\be
P( \Delta\vect{\theta} | s_i, \vect{\theta}_{\mathrm{true}} ) \propto P^{\mathrm{sys}}( \vect{\theta}_{\mathrm{true}} + \Delta\vect{\theta}, s_i ).
\ee
This result makes sense intuitively --- our potentially biased inferred posterior $P^{\mathrm{sys}}( \vect{\theta}, s_i )$ can be interpreted as the probability that the systematic bias $\Delta\vect{\theta}$ is equal to $\vect{\theta} - \vect{\theta}_{\mathrm{true}}$. Thus if we find $P^{\mathrm{sys}}( \vect{\theta}, s_i )$ is consistent with $\vect{\theta} = \vect{\theta}_{\mathrm{true}}$, this implies $\Delta\vect{\theta}$ is consistent with zero.

Assuming our $N$ simulated realizations are independent\footnote{We note that for neither of the simulation suites used here is this really true. For BCC, each set of 6 DES Y1 realizations is sourced from the same set of three N-body simulations, while for MICE, the two realizations are sourced from the same N-body simulation. While we ensure our Y1 realizations are extracted from non-overlapping regions, there will still be large-scale correlations between them. For our application, ignoring this is conservative, since unaccounted-for correlations between the realizations would tend to lead to fluctuations in the inferred parameters from their true values that are correlated between realizations, leading to over-estimates of systematic biases.}, it follows that 
\be
P( \{s_i\} |  \vect{\theta}_{\mathrm{true}}, \Delta\vect{\theta} ) = \prod_{i=1}^{N} P( s_i |  \vect{\theta}_{\mathrm{true}}, \Delta\vect{\theta} ),
\ee
and
\be
P( \Delta\vect{\theta} | \{s_i\}, \vect{\theta}_{\mathrm{true}} ) \propto \prod_{i=1}^{N} P^{\mathrm{sys}}( \vect{\theta}_{\mathrm{true}} + \Delta\vect{\theta}, s_i ).
\label{eq:multisim}
\ee
In summary, we can estimate the systematic bias in our inferred parameters by computing the (potentially biased) parameter posterior $P^{\mathrm{sys}}(\vect{\theta},s_i)$ from each simulation realization, and taking their product. 

\begin{figure*}
\includegraphics[width=0.9\columnwidth,trim={0.cm 0. 1.5 1.5},clip]{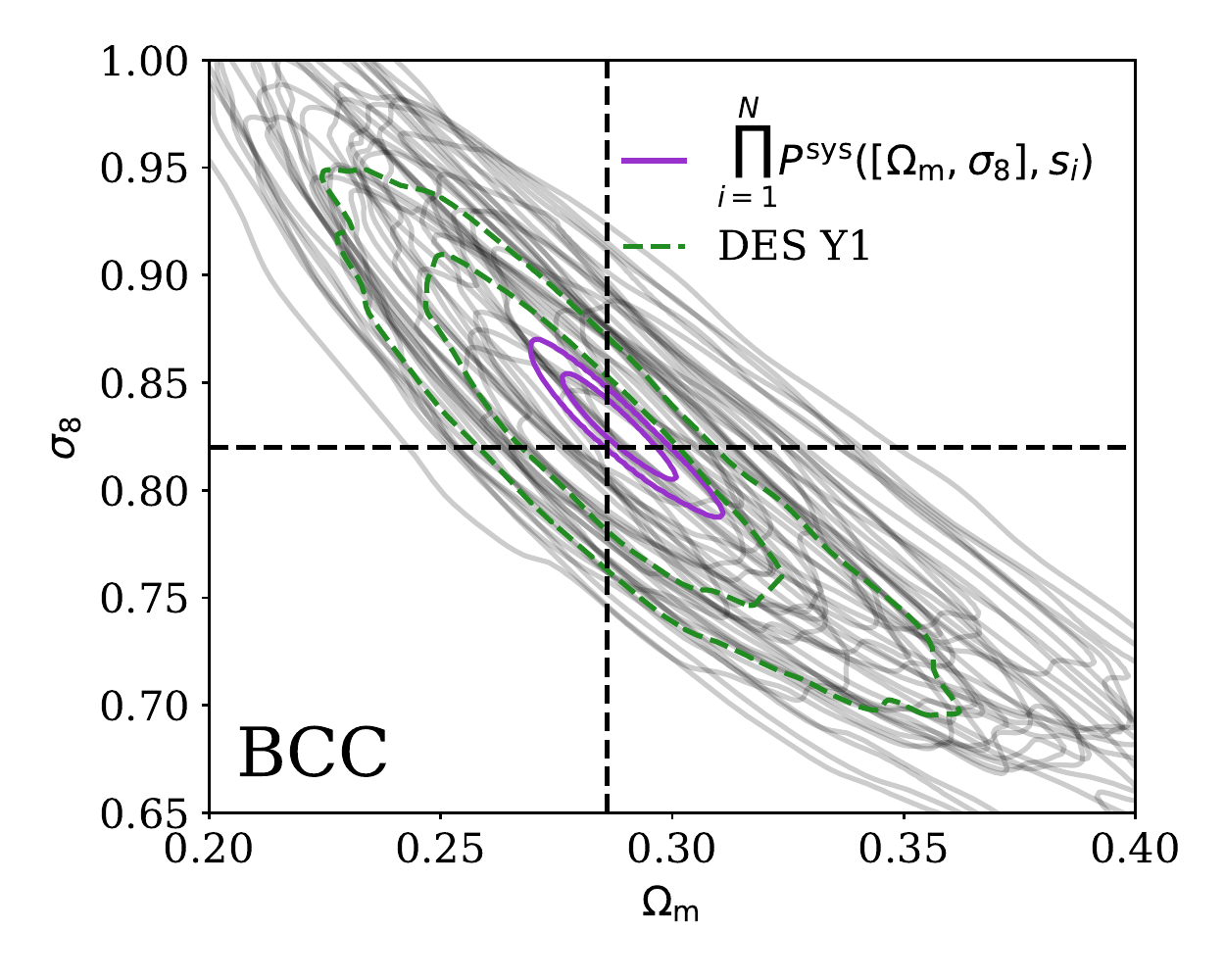}
\includegraphics[width=0.9\columnwidth,trim={0.cm 0. 1.5 1.5},clip]{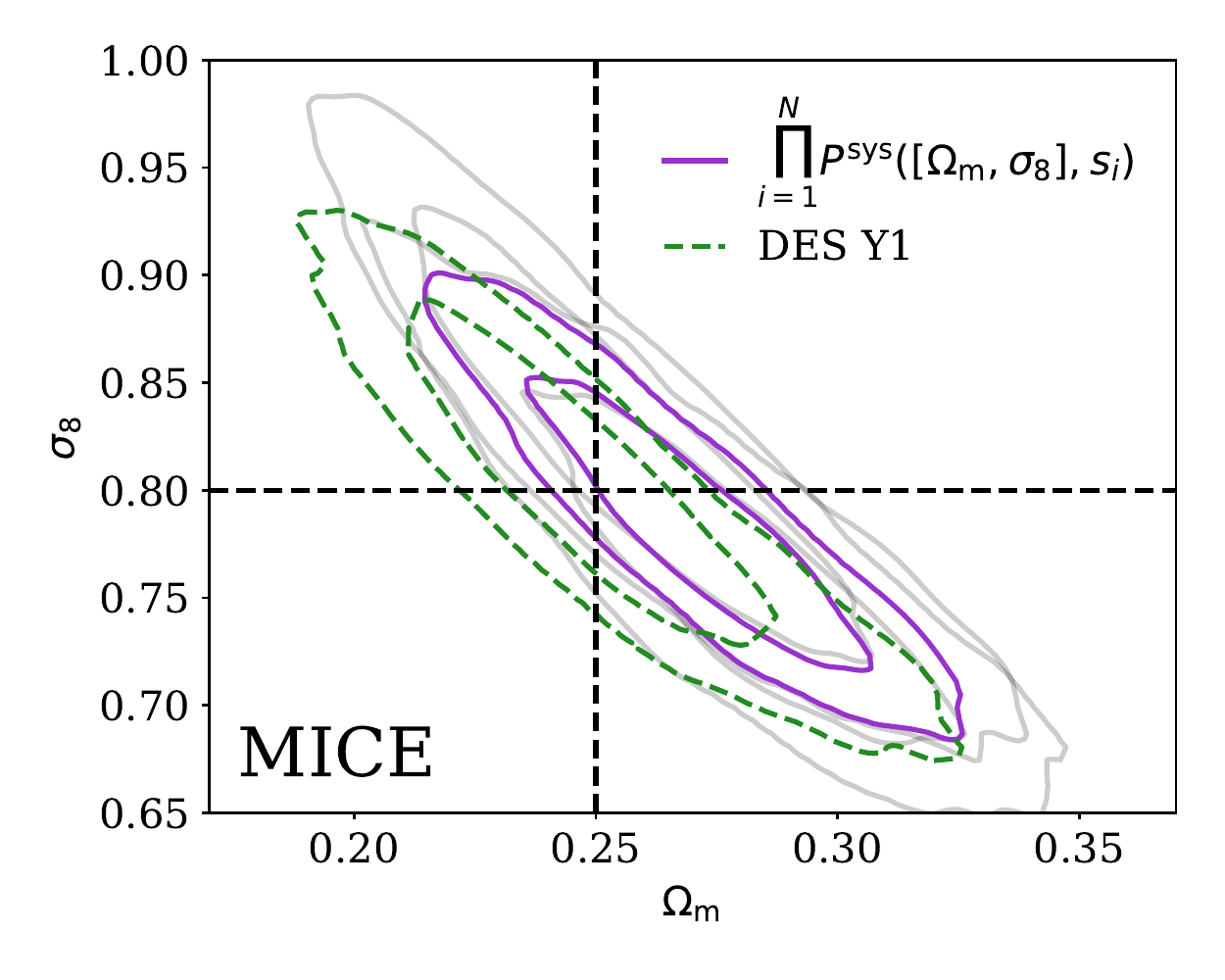}
\includegraphics[width=0.9\columnwidth,trim={0.cm 0. 1.5 1.5},clip]{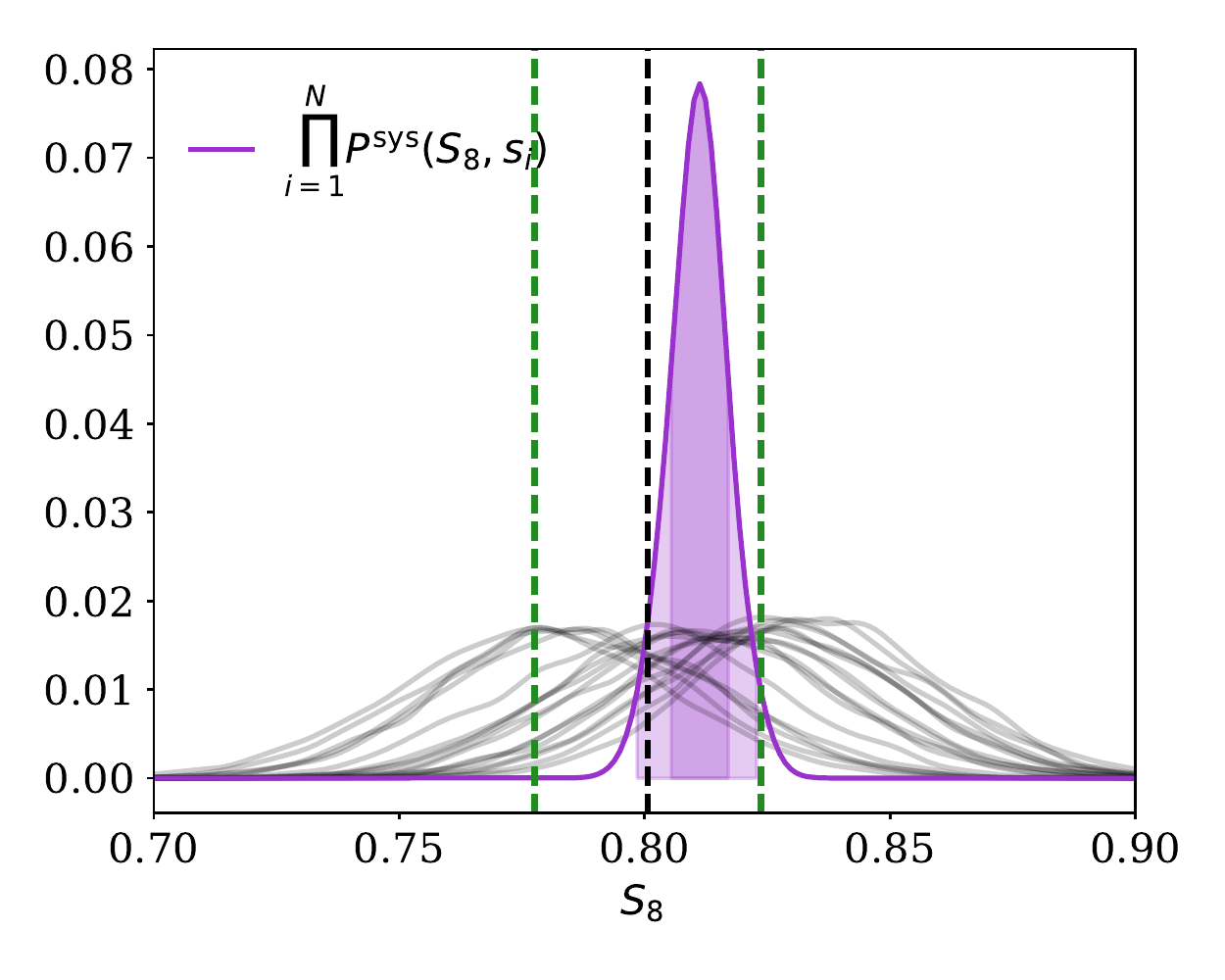}
\includegraphics[width=0.9\columnwidth,trim={0.cm 0. 1.5 1.5},clip]{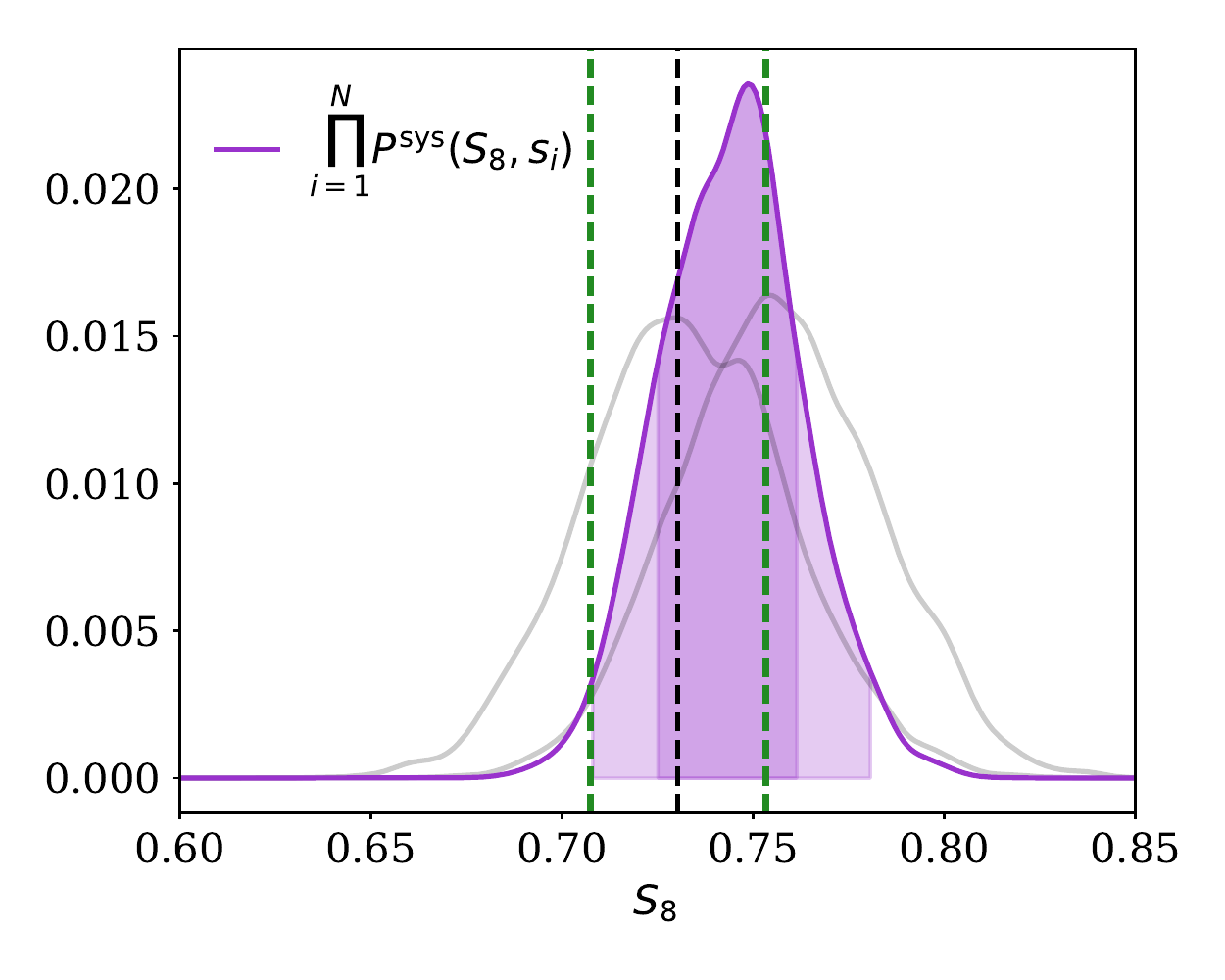}
\includegraphics[width=0.9\columnwidth,trim={0.cm 0. 1.5 1.5},clip]{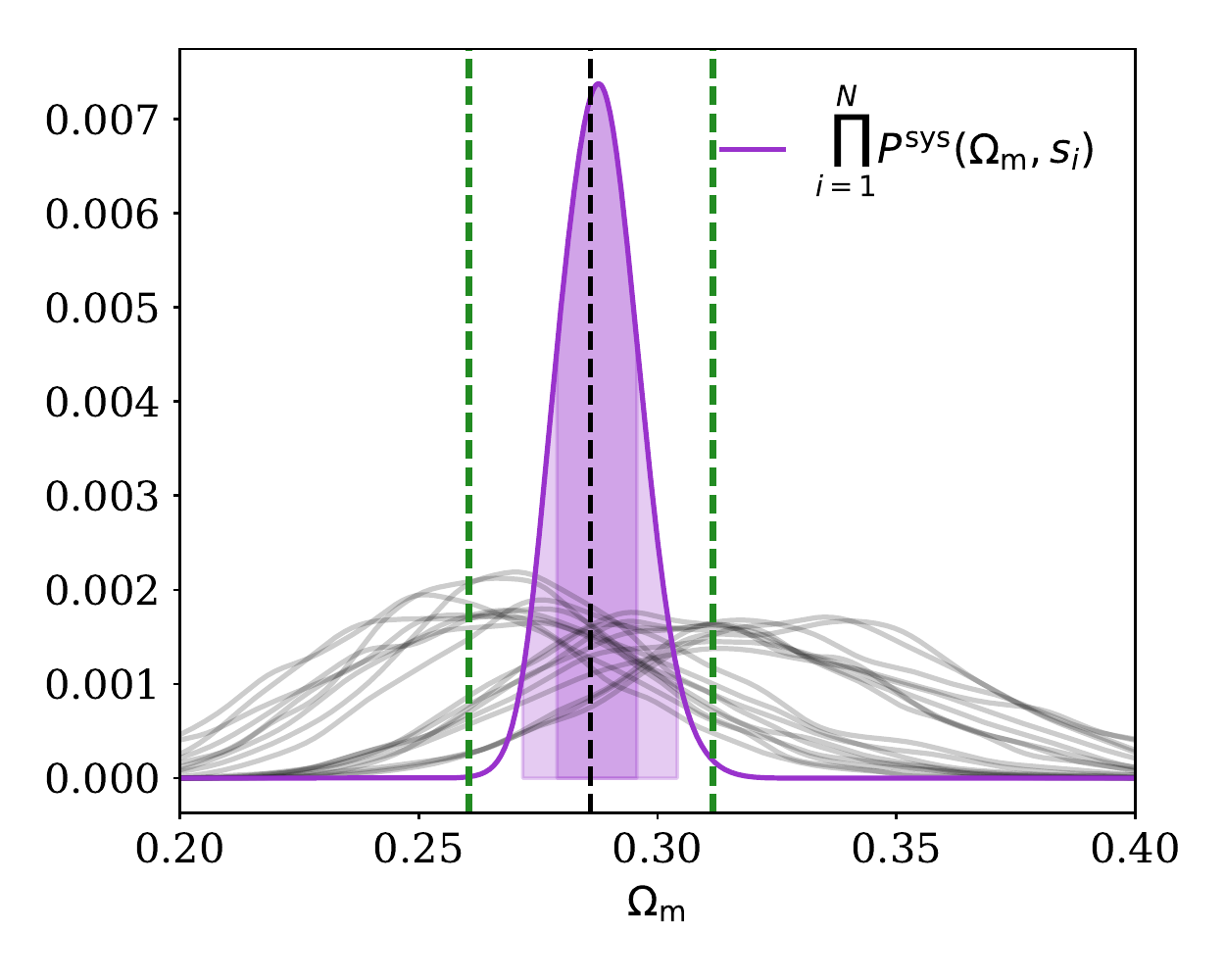}
\includegraphics[width=0.9\columnwidth,trim={0.cm 0. 1.5 1.5},clip]{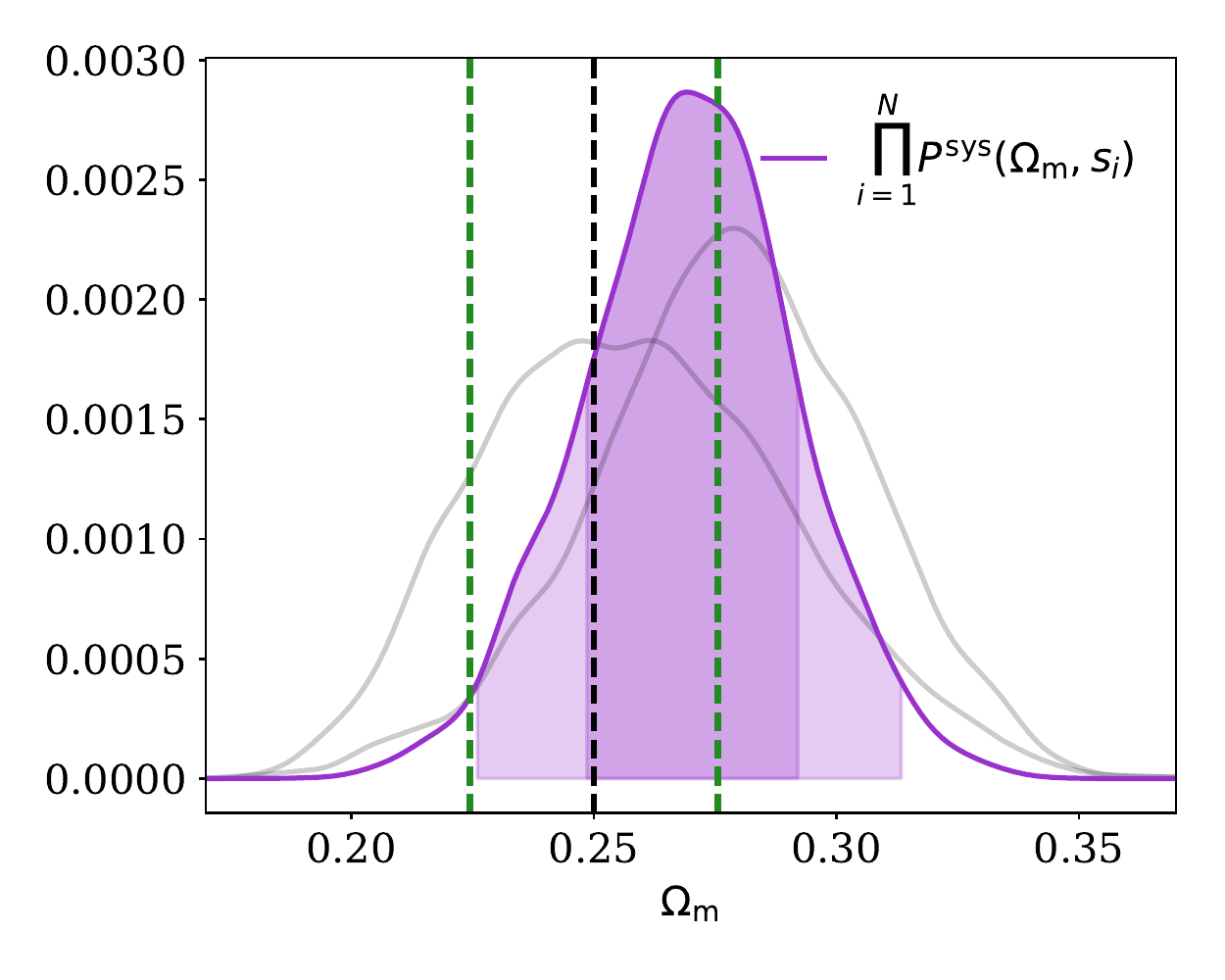}
\caption[]{
Cosmological constraints from all three two-point functions for the BCC ({\em left panels}, 18 realizations) and MICE ({\em right panels}, 2 realizations) simulation suites. The top panels show constraints on the present day matter density, \om\, and the clustering amplitude, \sig. Contours contain 68\% and 95\% of the posterior probability. Grey contours show constraints from individual simulation realizations, while purple contours show the combination of these posteriors (see \eqn{eq:multisim}). The middle and bottom panels show the marginalized constraints on $S_8 = \sig \times (\om/0.3)^{0.5}$, and \om\ respectively. In all panels, the true parameter values (i.e. those input to the simulations) are indicated by the black dashed lines. In the top panels, the green dashed lines indicate the 68\% and 95\% confidence regions  recovered from the real DES Y1 data in \citet{keypaper}, shifted to be centered on the input cosmology to the simulations. In the middle and lower-panels, the green dashed lines indicate the size of the $1-\sigma$ uncertainty from \citet{keypaper}.
}
\label{fig:cosmo1}
\end{figure*}

In this section we focus on studying biases in \om\ and \sig, the only two cosmological parameters well-constrained by DES Y1 data in \citet{keypaper}.
The top panels of \fig{fig:cosmo1} shows constraints on \om\ and \sig\ from the BCC ({\em top-left}) and MICE ({\em top-right}) simulation suites, using all three two-point functions ($\xipm(\theta)$, $\gamma_t(\theta)$ and $w(\theta)$). The dark orchid contours are the combined constraints from all realizations, calculated from the single-realization posteriors (shown in grey), using \eqn{eq:multisim}. Here and in all other plots, the contours indicate the 68\% and 95\% confidence regions.

For both MICE and BCC, the true cosmology (indicated by the black dashed lines) is within the 95\% contour, so we find no strong evidence for a non-zero $\Delta\vect{\theta}$.
In the middle panels, we show the marginalized posteriors for the well-constrained parameter combination $S_8 = \sig(\om/0.3)^{0.5}$; again, the true value of $S_8$ is within the 95\% confidence region (indicated by the lighter shaded region under the posterior curve). Finally, the lower panels show the marginalized posteriors for \om, which again are fully consistent with the true value for both BCC and MICE.

For comparison, in all panels we also indicate with green dashed lines the uncertainty on the parameters recovered from the real DES Y1 data in \citet{keypaper}, as $68\%$ and $95\%$ marginalised contours in the top row, and marginalised $1\sigma$ uncertainties in the middle and bottom rows. These uncertainties include marginalisation over nuisance parameters, including those accounting for shear calibration uncertainty and intrinsic alignments, which were not considered in the analysis of the simulations in this work.

\begin{figure*}
\includegraphics[width=0.9\columnwidth,trim={0.cm 0. 1.5 1.5},clip]{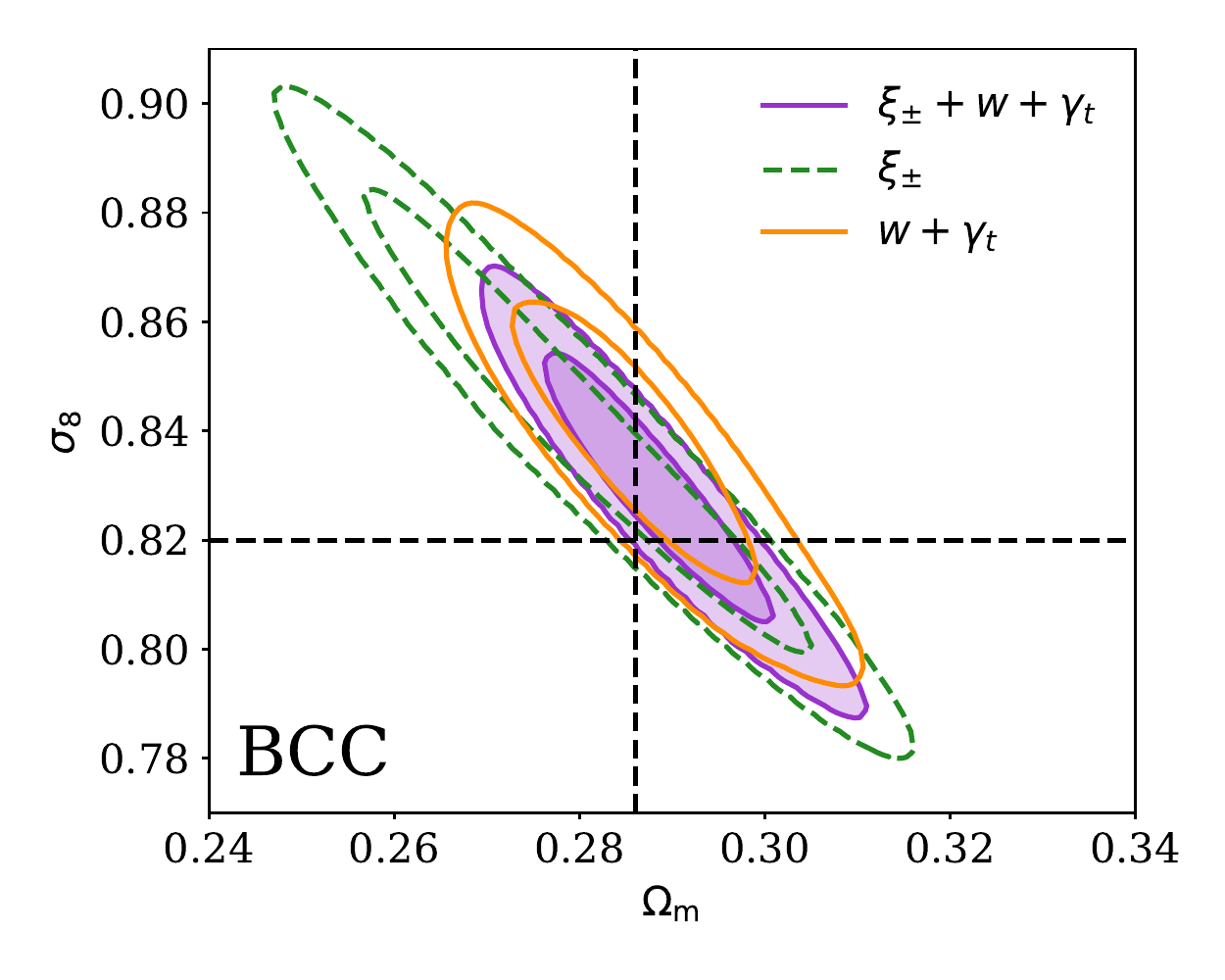}
\includegraphics[width=0.9\columnwidth,trim={0.cm 0. 1.5 1.5},clip]{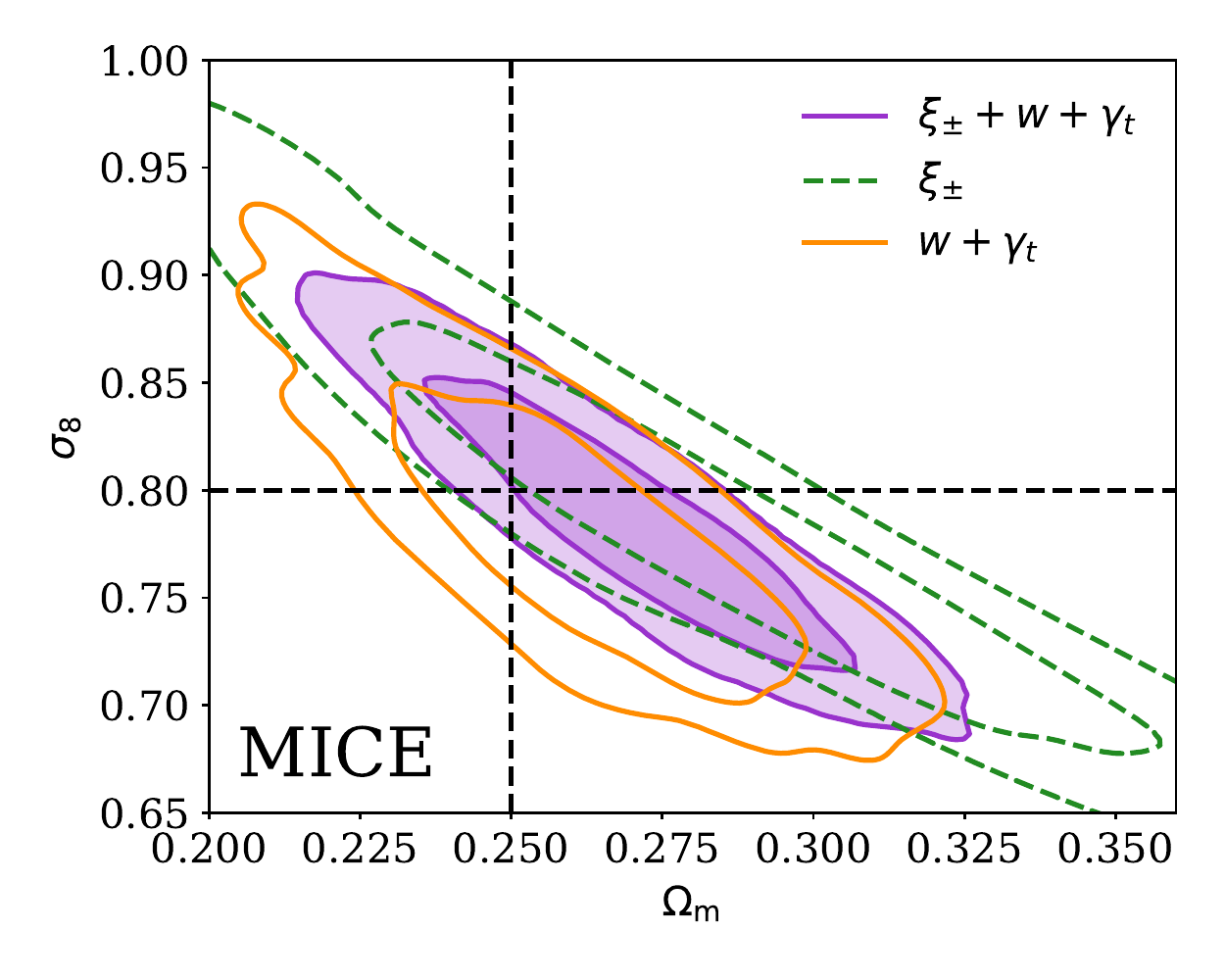}
\caption[]{
Constraints on the present day matter density, \om\, and the clustering amplitude, \sig\ from all three two-point functions (filled purple contours, labeled `$\xipm+\gamma_t+w$'), cosmic shear-only  (dashed green contours,  labeled `$\xipm$'), and galaxy-galaxy lensing and clustering (solid-lined orange contours, labeled `$\gamma_t+w$'), for the BCC simulations (left panel) and MICE simulations (right panel). We show combined constraints from all simulation realizations. The intersection of the black dashed lines indicates the true parameter values (i.e. those input to the simulations).
}
\label{fig:cosmo2}
\end{figure*}

\fig{fig:cosmo2} meanwhile shows the constraints in the $\om-\sig$ plane for subsets of the datavector for the BCC ({\em left panel}) and MICE simulations ({\em right panel}). Again, these contours represent the combination of the posteriors from all  individual simulation realizations. In both panels, the constraints from cosmic-shear only are shown as the green dashed unfilled contours (labeled `$\xipm$'), those from galaxy--galaxy lensing and galaxy clustering are shown as solid orange unfilled contours (labeled `$w+\gamma_t$'), and those from all three two-point functions are shown as filled dark-orchid contours (labeled `$\xipm+w+\gamma_t$', these are the same as those in the upper panels of \fig{fig:cosmo1}). For both BCC and MICE, we see good agreement between the datavector subsets, and no 
evidence for disagreement with the true cosmology.

We can use these parameter constraints to make estimates of the biases in inferred parameters produced by systematic biases in the parameter inference (assuming perfect simulations). Note that these estimates are conditional on the specific set of simulated data that were realized, $\{s_i\}$. Clearly it is desirable to have systematic biases be sub-dominant to statistical uncertainties. Therefore, for constraints from all three two-point functions, we report the probability that the bias in the inferred parameter $\theta$ (where $\theta$ is \om\ or $S_8$) are within $\sigma_{Y1}(\theta)$ of their true values, where $\sigma_{Y1}(\theta)$ is the $1\sigma$ uncertainty on parameter $\theta$ recovered by \citet{keypaper}. 
We denote this quantity $P(\Delta\theta < \sigma_{Y1})$. \citet{keypaper} find $\sigma_{Y1}(S_8)=0.023$ and $\sigma_{Y1}(\om)=0.026$ (for convenience, we use half the difference between the reported upper and lower 68\% confidence limits rather than propagating asymmetric errorbars).
Of course this should not be interpreted as an estimate of the impact of \emph{all} systematics errors, only those tested by the simulations.
With this caveat duly noted, $P(\Delta\theta < \sigma_{Y1})$ can be calculated as the integral of the posteriors in the middle and lower panels of \fig{fig:cosmo1} between the dashed lines, for $\theta=S_8$ and \om\ respectively. Ideally one could also calculate e.g. $P\left( \Delta\theta < \frac{1}{2} \sigma_{Y1} \right)$, however, given our available simulation volume, we do have sufficient statistical power to meaningfully constrain parameter biases to this precision.

For the BCC simulations, we find $P(\Delta S_8 < \sigma_{Y1}) = 0.98$ and $P(\Delta \om < \sigma_{Y1}) = 1.00$ (we report these probabilities to two decimal places), indicating that we can be confident that systematic biases in our inference of $S_8$ and $\om$ from the BCC simulations are less than the DES Y1 $1\sigma$ uncertainty for those parameters.

For the MICE simulations, we find $P(\Delta S_8 < \sigma_{Y1}) = 0.66$ and 
$P(\Delta \om < \sigma_{Y1}) =0.57$. 
Again, more simulation volume is required to make strong statements about the sub-dominance (or not) of systematic errors to statistical errors from the MICE simulations. This does not make the analysis of the MICE simulations a pointless exercise; we can comfort ourselves with the fact that we could have uncovered large (i.e. larger than the DES Y1 $~1\sigma$ uncertainty) systematic biases, and did not.

\tab{tab:biasconstraints} summarizes our parameter bias results for both simulation suites and all subsets of the datavector considered.

\begin{table*}
\caption{A summary of constraints on parameter biases inferred from both simulation suites. The $\Delta\om$ and $\Delta S_8$ provide the absolute bias in $\om$ and $S_8$ (and 68\% confidence intervals) with respect to the truth input to the simulations. For comparison, the uncertainty on these parameters for the accompanying analysis in \citet{keypaper} is 0.026 and 0.023 respectively; statistically significant biases of this level would be a cause for concern.}
\begin{tabular}{ccccccccc}
\hline
Dataset & $\Delta\om$  & $P(\Delta\om < \sigma_{Y1})$ & $\Delta S_8$ & $P(\Delta S_8 < \sigma_{Y1} )$ \\
\hline
BCC $\xipm+w+\gamma_t$ & $ 0.0017 \pm 0.0084 $ & $ 1.00 $ & $ 0.0106 \pm 0.0058 $ & $ 0.97 $ \\
\hline
BCC $\xipm$ & $ -0.0125 \pm 0.0120 $ & $ 0.84 $ & $ 0.0067 \pm 0.0059 $ & $ 0.99 $ \\
\hline
BCC $w+\gamma_t$ & $ -0.0010 \pm 0.0085 $ & $ 1.00 $ & $ 0.0156 \pm 0.0077 $ & $ 0.81 $ \\
\hline
MICE $\xipm+w+\gamma_t$ & $ 0.0191 \pm 0.0217 $ & $ 0.57 $ & $ 0.0183 \pm 0.0182 $ & $ 0.67 $ \\
\hline
MICE $\xipm$ & $ 0.0198 \pm 0.0434 $ & $ 0.39 $ & $ 0.0239 \pm 0.0213 $ & $ 0.49 $ \\
\hline
MICE $w+\gamma_t$ & $ 0.0052 \pm 0.0226 $ & $ 0.68 $ & $ -0.0024 \pm 0.0262 $ & $ 0.61 $ \\
\hline
\label{tab:biasconstraints}
\end{tabular}
\end{table*}

\subsection{Photometric redshifts}\label{sec:photozs}

\begin{figure}
\includegraphics[width=0.9\columnwidth]{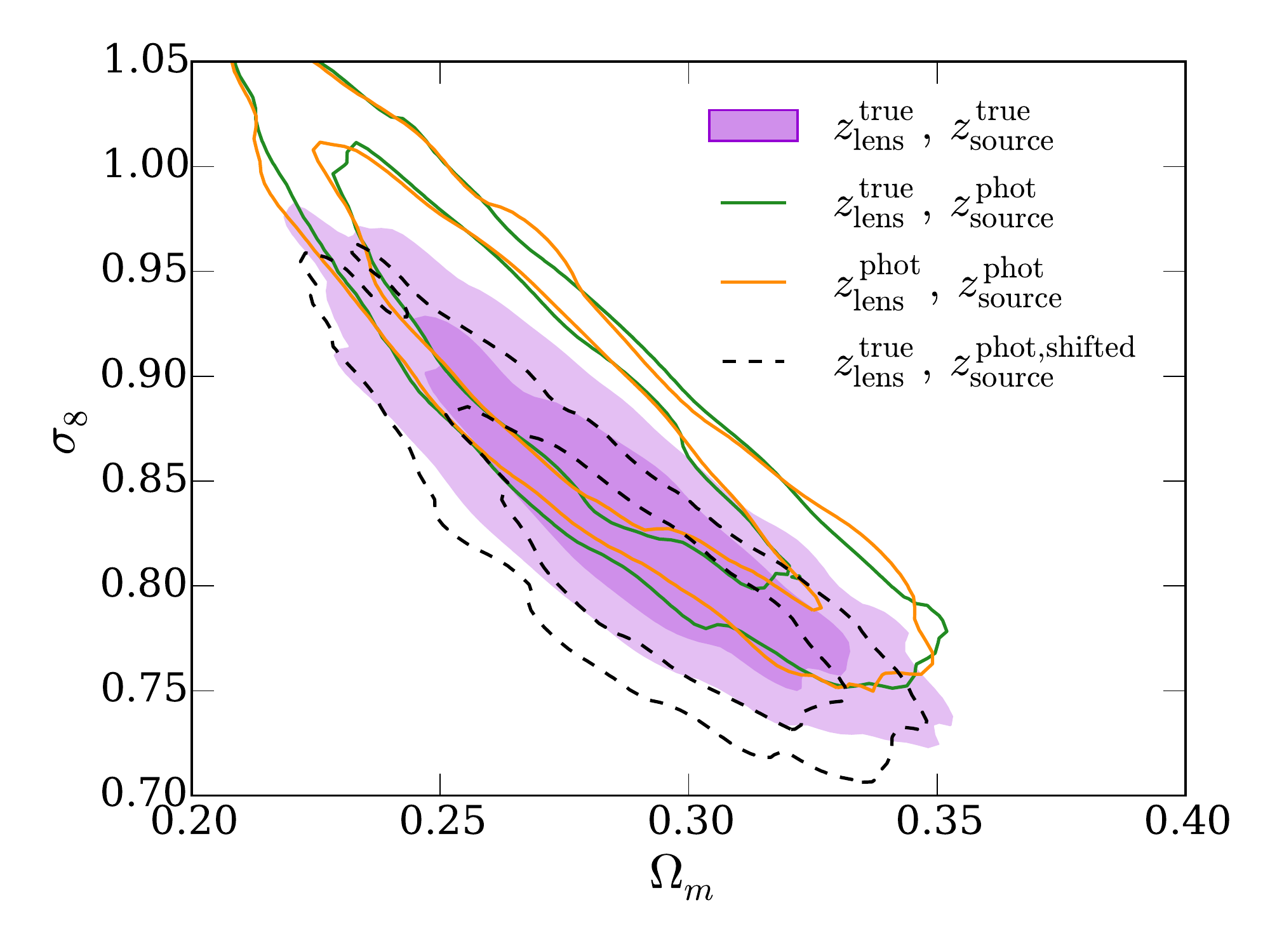}
\caption[]{The impact of using photometric redshift estimates. For all contours, the datavector is the mean of all BCC realisations. Purple filled contours use the true redshift distributions for both lens and source galaxies. Note in this case galaxies are binned according to their \photoz, but the $n(z)$ for each redshift bin is estimated using true redshifts. The green unfilled contours use source redshift distributions estimated using BPZ \photoz s, while the orange unfilled contours additionally use the photometric redshift estimates from the the \redmagic\ algorithm, \zred\ for the lenses. The black dashed contours use BPZ to estimate the source $n(z)$s, shifted in redshift to have the correct mean (see \sect{sec:photozs} for details).}
\label{fig:pz}
\end{figure}

Photometric redshift estimation is one of the major challenges for extracting precise cosmological information from imaging surveys (e.g. \citealt{schmidt14}). In this section we test the effects of photometric redshift biases on the inferred cosmological parameters for the BCC simulation suite. For the fiducial analyses of real DES Y1 data, BPZ was used to estimate the $n(z)$s of the source sample, as described in \citet{photoz}. These $n(z)$ estimates were further refined by comparison to two independent \photoz\ methods, and shifts of the form $n^i(z) \rightarrow n^i(z+\delta z^i)$ were applied to each redshift bin $i$, with uncertainty in the $\delta z^i$ marginalised over as part of the cosmological parameter estimation (with Gaussian priors of width $\left[0.016, 0.013, 0.011, 0.022\right]$).

We do not implement the two independent \photoz\ methods used to correct the BPZ $n(z)$s on these simulations (this would require significantly expanding the scope of the simulations), but as an idealised proxy, we do apply shifts $\delta z^i$ to the BPZ $n(z)$ estimates such that they have the correct mean redshift (see case (iii) below).  

For the BCC simulations (we do not use \photoz\ estimates for the source sample in MICE), we compare the recovered cosmological parameters in the following cases:
\begin{enumerate}
\item{We use the true redshifts to construct the $n(z)$s for both lenses and sources; this was our fiducial treatment in \sect{sec:cosmo}}.
\item{We use true redshifts to construct the $n(z)$ for the lenses, but BPZ estimates to construct the $n(z)$ for the sources. For this case we marginalize over a shift $\delta z^i$ for each source redshift bin $i$, with independent Gaussian priors with mean zero, and width 0.02 (this is same order as those used in \citealt{keypaper})}
\item{The same as case (ii), but we first shift the BPZ $n(z)$ estimates such that they have the correct mean redshift.}
\item{The same as case (ii), but we now also use photometric redshifts to construct the lens sample $n(z)$s.}
\end{enumerate}

Given the size of the DES Y1 area, there is very little variation in the $n(z)$s between simulated realizations, therefore we can assess the effect of using photometric redshift biases by comparing the cosmological constraints inferred from the mean datavector across all realizations when we use the true redshifts to construct the theoretical prediction for the datavector, to when we use photometric redshift estimates in the theoretical prediction. We use a covariance matrix appropriate for a single DES Y1 realization in the likelihood calculation, which naturally puts any differences in the contours in the context of DES Y1 uncertainties.

\fig{fig:pz} demonstrates that the inferred cosmological parameters do change significantly when using photometric redshifts in the BCC simulations. The green contour illustrates case (ii); in this case $S_8$ is significantly biased with respect to the fiducial result (the filled purple contour) by 0.038, greater than the achieved $1\sigma$ uncertainty for DES Y1. 
The dashed black contour shows case (iii), where the BPZ $n(z)$ estimates are first shifted to have the correct mean. The result is improved, but the bias in $S_8$ of 0.020 with respect to the fiducial case is still non-negligible. Using photometric redshifts for the lens $n(z)$s does not introduce significant parameter bias - the $S_8$ bias for case (iv) differs from that for case (ii) by only $4\times10^{-3}$. This implies that the \zred\ photo-zs in BCC are comfortably performing sufficiently well for DES Y1.

\begin{figure}
\includegraphics[width=0.9\columnwidth,trim={0.cm 0. 1.5 1.5},clip]{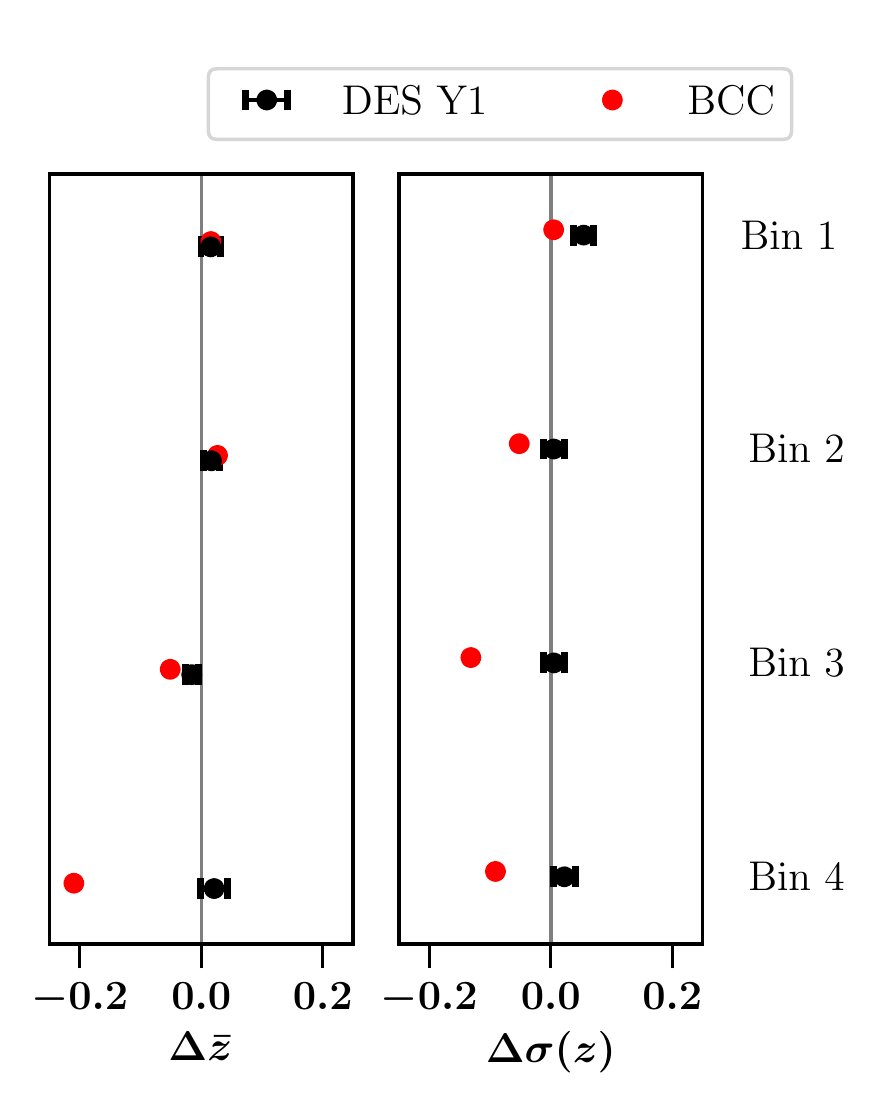}
\caption[]{
Estimates of the bias in the mean redshift ($\bar{z}$) and the bias in the redshift standard deviation ($\sigma(z)$) reported by BPZ for each source redshift bin. For the BCC simulations (red points) these biases can be computed exactly, while for the real DES Y1 data (black points), these are estimated from comparison with a re-weighted COSMOS sample with high-precision redshift information \citep{photoz}. For the two highest source redshift bins, we see much larger biases in both the mean redshift and the redshift standard deviation for BCC than we do for the DES Y1 data. While biases in $\bar{z}$ can be accounted for using the $\delta z^i$ parameterization described in \sect{sec:photozs}, biases in $\sigma(z)$ are not accounted for by this  parameterization. 
}
\label{fig:nz_bias}
\end{figure}

It appears then that marginalizing over $\delta z_i$ for the source redshift bins with the above priors is an insufficiently flexible scheme to account for biases in the BPZ $n(z)$ estimates for the BCC simulations. On the real DES data, the opposite conclusion was reached in \citet{photoz}, with BPZ performing sufficiently well in comparison to a re-weighted COSMOS sample with high-precision \photoz s \citep{laigle16} and cross-correlation methods. Given this result, it is likely that BPZ is performing worse on the BCC simulations than the real DES data. This is demonstrated in \fig{fig:nz_bias}, where we show estimates of the bias in the mean redshift ($\bar{z}$) and the bias in the width (the standard deviation, $\sigma(z)$) of each source redshift bin. For the BCC $n(z)$'s, we can use true redshift information to calculate these quantities exactly, while for the DES Y1 data, these biases are estimated by comparison to the aforementioned COSMOS sample. For the DES Y1 points, error bars indicate the uncertainty of the COSMOS-based estimates of $\bar{z}$ and $\sigma(z)$, using the methodology of and including all effects discussed in \citet{photoz}.

Particularly for the highest two redshift bins, for BCC we see large biases in both the mean redshift, and the width of the redshift distribution. While the impact of the former could be mitigated by the $\delta z_i$ nuisance parameters, biases in the width of the $n(z)$s can not be. This is the likely explanation for why marginalizing over $\delta z_i$ did not mitigate BPZ redshift biases sufficiently for BCC.

The conclusion that the BPZ performance is worse on BCC than on the real DES data of course depends on the reliability of the COSMOS-based \photoz\ validation, especially for the highest-redshift bin where a clustering-based n(z) estimate was not available as additional validation. Several potential sources of biases in the n(z) estimation using the COSMOS \photoz s are investigated and quantified in \citet{photoz}; these contribute to the error bars on the DES Y1 points in \fig{fig:nz_bias}, which are still much smaller than the differences apparent between DES Y1 and BCC. As discussed in \citet{photoz}, for the DES Y1 magnitude range ($\lesssim 23.5$), $n(z)$ biases due to errors in the COSMOS \photoz s of greater than a few percent seem very unlikely based on the results of \citet{laigle16}. We find it very unlikely therefore that biases at the level we see for BPZ on BCC (e.g. 20\% in mean redshift for the highest z bin) could be present for the BPZ and COSMOS estimates on the DES Y1 data.

Potential reasons for the poor performance of BPZ will be explored further in \citet{derose2018}. Particularly at high redshift, there may be a mismatch between the BPZ templates and the galaxy colors simulated in the BCC since the BPZ templates include a redshift evolution correction based on higher redshift spectroscopic data (see \citealt{photoz}) that is not present in the low redshift SDSS data used by ADDGALS (see \sect{sec:addgals}). As discussed in \sect{sec:wlsample}, we note also that our procedure for selecting the source sample from the BCC was highly simplified compared to the procedure on the data. Even if the galaxy colors in the BCC matched the real Universe perfectly, this difference in selection could produce a source sample in BCC on which BPZ performs differently as compared to the real data.

We note here that generating mock galaxies with realistic joint distributions of clustering properties, colors and redshift down to the magnitude limits, and in the redshift ranges required for DES, is extremely challenging.
Iterative improvements in empirical and theoretical galaxy models with comparison to DES and other large photometric and spectroscopic datasets will likely be required to meet this challenge. 

\section{Discussion}\label{sec:discussion}

\begin{table*}
\caption{A list of phenomena relevant to the combined clustering and weak lensing analysis presented here that exist in real data. We indicate whether they are included in the simulations used in this work, and thus whether our treatment of the effect is tested by the analysis of the simulations presented here. 
`Higher-order lensing effects' refers to contributions to the weak lensing shear at higher than first order in the gravitational potential (see e.g. \citealt{krause10}) that are captured by full ray-tracing as in the BCC. `Beyond Limber angular statistics + RSD' refers to the deviation of the angular power spectra in the simulations from that predicted assuming the Limber approximation and ignoring redshift space distortions, as in our modeling.
Although the BCC catalogs do provide photo-z estimates, the differences between the performance of BPZ on the simulated and real data limited their usefulness (see \sect{sec:photozs}) --- the `$(\checkmark)$' symbol in the `Photo-z bias' row reflects this.}
\begin{tabular}{ccccccccc}
\hline
\textbf{Real-data complexity} & \textbf{Included in BCC} & \textbf{Included in MICE} \\
\hline
Photometry and shear estimation biases & $\times$ & $\times$ \\ 
\hline
Higher-order lensing effects & \checkmark & $\times$ \\ 
\hline
Intrinisc Alignments & $\times$ & $\times$  \\ 
\hline
Nonlinear galaxy bias & \checkmark & \checkmark \\ 
\hline
Beyond-Limber angular statistics + RSD & \checkmark & \checkmark \\ 
\hline
Magnification effect on number counts & \checkmark & \checkmark \\
\hline
Baryonic effects on the matter distribution & $\times$ & $\times$ \\ 
\hline
Photo-z biases & $(\checkmark)$ & $\times$ \\ 
\hline
Spatially varying photometric noise & \checkmark  & \checkmark  \\ 
\hline
Non-Gaussian distributed datavectors & \checkmark & \checkmark \\ 
\hline
\label{tab:complexities}
\end{tabular}
\end{table*}


Combined weak lensing and clustering analyses on Stage III galaxy surveys are still in their infancy, but already present a significant step forward for cosmological inference from these surveys \citep{vanuitert17,keypaper,joudaki18}. Even in the absence of systematics, this multi-probe approach allows more cosmological information to be extracted, and when uncertain contributions from systematic effects are included, can greatly reduce the degradation in cosmological constraining power they cause. However, such analyses rely on various theoretical assumptions, and assume that observational and astrophysical systematics can be treated accurately, making validation of the methodology extremely important. 

In this work, we have attempted to validate the methodology used on DES Y1 data by performing a similar analysis of tailored survey simulations. Simulations have of course been extensively used in cosmology analyses of photometric datasets. Related recent cosmology analyses include  \citet{mandelbaum13}, who performed a detailed study of the galaxy--matter cross-correlation in simulations to validate their theoretical modeling. \citet{vanuitert17} presented a similar combined weak lensing and galaxy clustering analysis on KiDS data to that considered here and tested aspects of their analysis on tailored simulations. However, these did not include a realistic lens sample to test the galaxy bias modeling. In this work we go further than previous analyses in attempting to simulate both galaxy clustering and lensing observables (rather than quantities that are not directly observable such as the galaxy--matter correlation function), estimated from galaxy samples that are selected using the same or at least approximate versions of the galaxy selection process used on the real data. By using observable quantities, we ensure that higher-order effects like magnification and reduced shear are included, so we are implicitly testing the impact of ignoring these effects in our theoretical modeling. 

These simulations contain many of the complexities of real data: spatially varying magnitude errors due to depth variations affect galaxy selection, the statistical connection between galaxies and matter is more complex than our simple theoretical models, and photometric redshift algorithms have been implemented (albeit further work on the galaxy colors is required to make this last aspect more informative). \tab{tab:complexities} summarizes  complexities present in real data that are relevant to a galaxy clustering and weak lensing analysis, and indicates whether they are included in the simulations used in this work. For those that are included, our analysis of the simulations constitutes a validation of their treatment in our cosmological parameter estimation pipeline.

As indicated in \tab{tab:complexities}, there are various potential systematic effects in real data that are not included in the simulations used here. For example, image simulations are likely required to test the accuracy of photometry and shear estimation pipelines (e.g. \citealt{great08handbook,great3handbook}). In order to produce image simulations with realistic distributions of galaxy properties (including clustering), it will be desirable to propagate the type of simulation used in this work to the image level (rather than just the mock catalogs used here). Some progress on this has already been made by \citet{chang2015}. 

We also note that neither simulation suite used here includes galaxy intrinsic alignments, which potentially contaminates the shear correlation functions and the galaxy--galaxy lensing signal. We believe including intrinsic alignments in future galaxy survey simulations should be prioritized, as it is one of the major systematics faced by weak lensing analyses \citep{troxel15,Joachimi2015}. Another primary systematic for cosmic shear analyses, the effect of baryonic physics on the matter power spectrum (see e.g. \citealt{white04, zhan04, sembolini11}), is also not included here. These effects, often termed \emph{astrophysical systematics}, clearly depend on additional physics to that implemented in the N-body simulations used here. 
Both of these effects are areas of active investigation using hydrodynamic simulations (e.g. \citealt{chisari18,chisari17}).

The galaxy bias, especially on small scales, of course also depends on the simulation implementation. Here again, hydrodynamic simulations are arguably closer to a first-principles approach to simulating galaxies than the empirical relations used in BCC and MICE. However they have a much higher computational cost, and much of the relevant physics still occurs on scales below the resolution limit of any current simulations that are large enough in volume to be applicable to cosmological analyses. 
It is likely that iterative comparison of the galaxy survey data such as the Dark Energy Survey are providing, simulations which use empirical relations to add galaxies like the ones used here, and hydrodynamic simulations will be required to inform us of the true behavior of galaxy bias. Nonetheless, we have shown that reasonable models for how galaxies trace the density field, implemented in the BCC and MICE simulations, show no strong evidence of bias in cosmological parameter inference, which should provide confidence in the robustness of the cosmological parameter constraints presented in the companion papers \citep{keypaper,shearcorr}. 

Given the amount of simulation volume currently available for the MICE simulations, we are unable to make a very definitive statement about the size of systematic biases with relation to the DES Y1 parameter uncertainties. For the BCC simulations on the other hand we do find convincing evidence that inferred parameter biases are smaller than the DES Y1 $1\sigma$ uncertainties. Updated versions of both simulation suites that will be used for upcoming DES cosmology analyses are already reasonably advanced in their development, and will provide larger simulation volumes, as well as implementing improved empirical relations between galaxy colors and clustering properties.

\section{Acknowledgements}

Thanks to Chris Hirata for useful discussions.

Funding for the DES Projects has been provided by the U.S. Department of Energy, the U.S. National Science Foundation, the Ministry of Science and Education of Spain, 
the Science and Technology Facilities Council of the United Kingdom, the Higher Education Funding Council for England, the National Center for Supercomputing 
Applications at the University of Illinois at Urbana-Champaign, the Kavli Institute of Cosmological Physics at the University of Chicago, 
the Center for Cosmology and Astro-Particle Physics at the Ohio State University,
the Mitchell Institute for Fundamental Physics and Astronomy at Texas A\&M University, Financiadora de Estudos e Projetos, 
Funda{\c c}{\~a}o Carlos Chagas Filho de Amparo {\`a} Pesquisa do Estado do Rio de Janeiro, Conselho Nacional de Desenvolvimento Cient{\'i}fico e Tecnol{\'o}gico and 
the Minist{\'e}rio da Ci{\^e}ncia, Tecnologia e Inova{\c c}{\~a}o, the Deutsche Forschungsgemeinschaft and the Collaborating Institutions in the Dark Energy Survey. 

The Collaborating Institutions are Argonne National Laboratory, the University of California at Santa Cruz, the University of Cambridge, Centro de Investigaciones Energ{\'e}ticas, 
Medioambientales y Tecnol{\'o}gicas-Madrid, the University of Chicago, University College London, the DES-Brazil Consortium, the University of Edinburgh, 
the Eidgen{\"o}ssische Technische Hochschule (ETH) Z{\"u}rich, 
Fermi National Accelerator Laboratory, the University of Illinois at Urbana-Champaign, the Institut de Ci{\`e}ncies de l'Espai (IEEC/CSIC), 
the Institut de F{\'i}sica d'Altes Energies, Lawrence Berkeley National Laboratory, the Ludwig-Maximilians Universit{\"a}t M{\"u}nchen and the associated Excellence Cluster Universe, 
the University of Michigan, the National Optical Astronomy Observatory, the University of Nottingham, The Ohio State University, the University of Pennsylvania, the University of Portsmouth, 
SLAC National Accelerator Laboratory, Stanford University, the University of Sussex, Texas A\&M University, and the OzDES Membership Consortium.

Based in part on observations at Cerro Tololo Inter-American Observatory, National Optical Astronomy Observatory, which is operated by the Association of 
Universities for Research in Astronomy (AURA) under a cooperative agreement with the National Science Foundation.

The DES data management system is supported by the National Science Foundation under Grant Numbers AST-1138766 and AST-1536171.
The DES participants from Spanish institutions are partially supported by MINECO under grants AYA2015-71825, ESP2015-66861, FPA2015-68048, SEV-2016-0588, SEV-2016-0597, and MDM-2015-0509, 
some of which include ERDF funds from the European Union. IFAE is partially funded by the CERCA program of the Generalitat de Catalunya.
Research leading to these results has received funding from the European Research
Council under the European Union's Seventh Framework Program (FP7/2007-2013) including ERC grant agreements 240672, 291329, and 306478.
We  acknowledge support from the Australian Research Council Centre of Excellence for All-sky Astrophysics (CAASTRO), through project number CE110001020, and the Brazilian Instituto Nacional de Ci\^encia
e Tecnologia (INCT) e-Universe (CNPq grant 465376/2014-2).

This manuscript has been authored by Fermi Research Alliance, LLC under Contract No. DE-AC02-07CH11359 with the U.S. Department of Energy, Office of Science, Office of High Energy Physics. The United States Government retains and the publisher, by accepting the article for publication, acknowledges that the United States Government retains a non-exclusive, paid-up, irrevocable, world-wide license to publish or reproduce the published form of this manuscript, or allow others to do so, for United States Government purposes.

This research used resources of the National Energy Research Scientific Computing Center, a DOE Office of Science User Facility supported by the Office of Science of the U.S. Department of Energy under Contract No. DE-AC02-05CH11231. Part of the BCC simulations used in this studies were performed using resources provided by the University of Chicago Research Computing Center, which we acknowledge for its support.




\bibliographystyle{mnras}
\bibliography{references} 



\appendix

\section{Two-point correlation function measurements}\label{app:tpt_meas}

\begin{figure*}
\includegraphics[width=14cm]{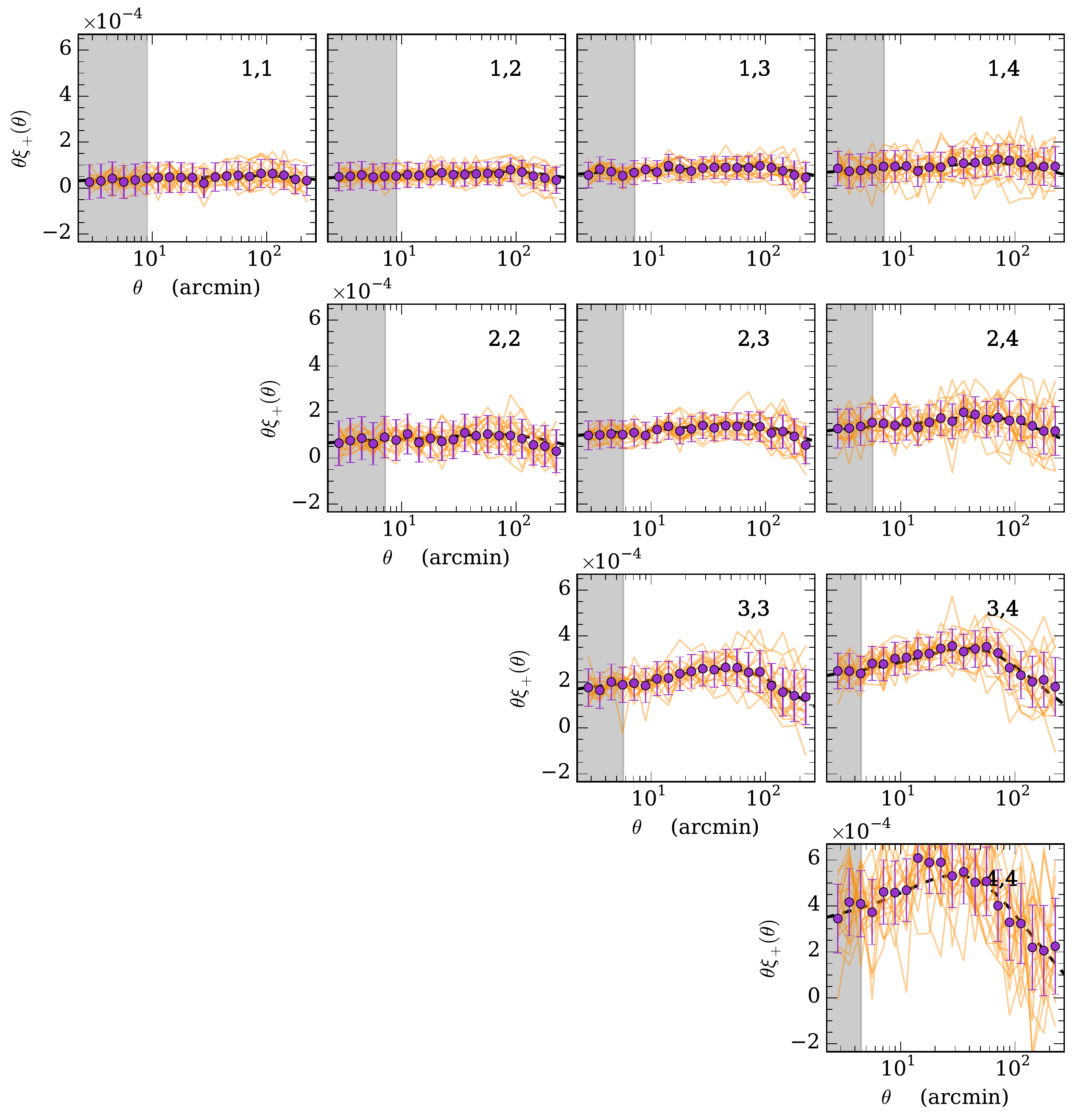}
\caption[]{Measurement of $\xip(\theta)$ in the BCC simulations. Dark orchid data points are the mean across all 18 simulations, while orange lines indicate measurements from individual realizations. Errorbars indicate the expected uncertainty from a single DES Y1 realization. The line is the theoretical prediction assuming the true cosmology. Grey shaded regions are excluded from the fiducial analysis.}
\label{fig:xip}
\end{figure*}

\begin{figure*}
\includegraphics[width=14cm]{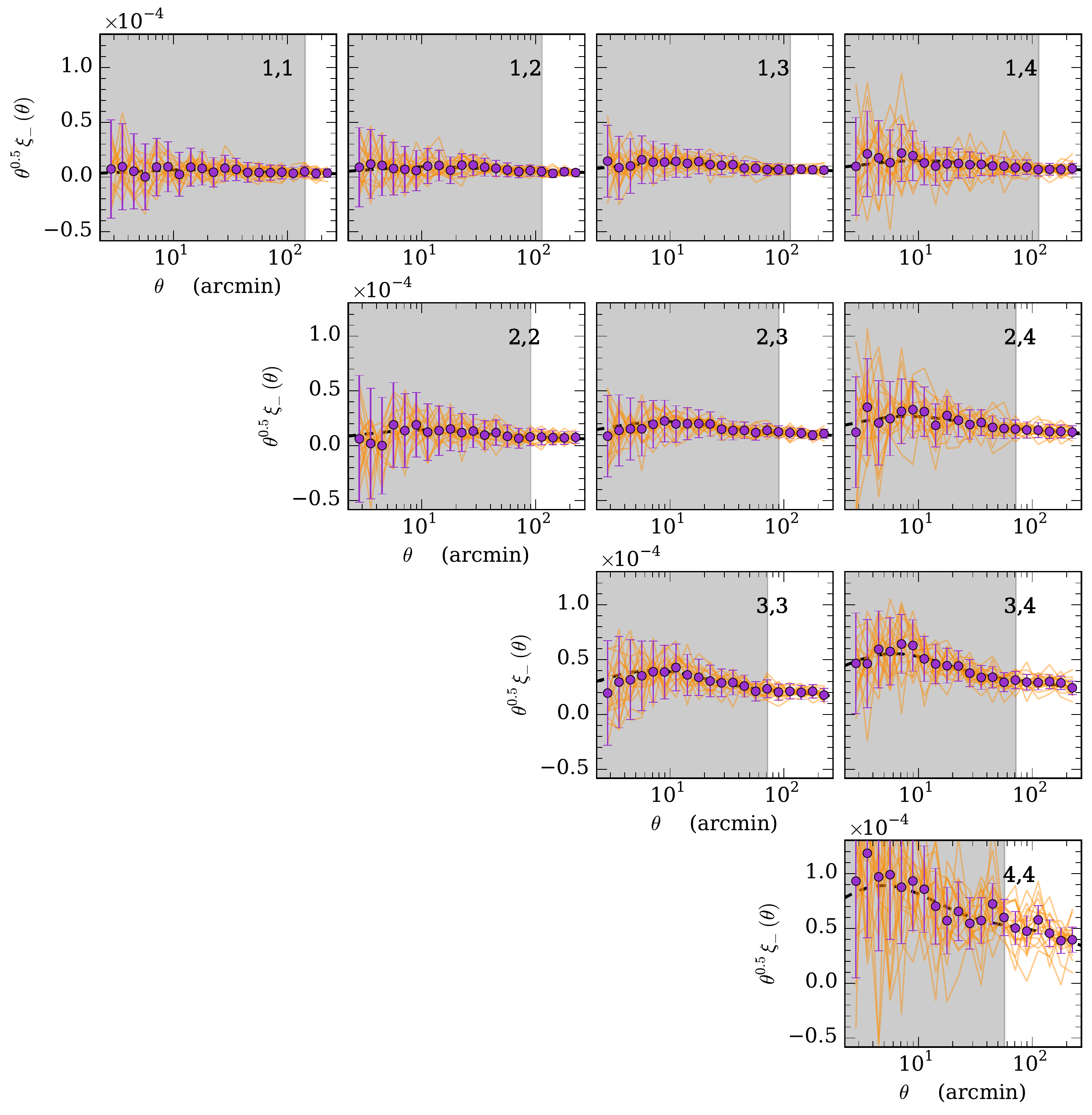}
\caption[]{Measurement of $\xim(\theta)$ in the BCC simulations. Dark orchid data points are the mean across all 18 simulations, while orange lines indicate measurements from individual realizations. Errorbars indicate the expected uncertainty from a single DES Y1 realization. The line is the theoretical prediction assuming the true cosmology. Grey shaded regions are excluded from the fiducial analysis.}
\label{fig:xim}
\end{figure*}

\begin{figure*}
\includegraphics[width=14cm]{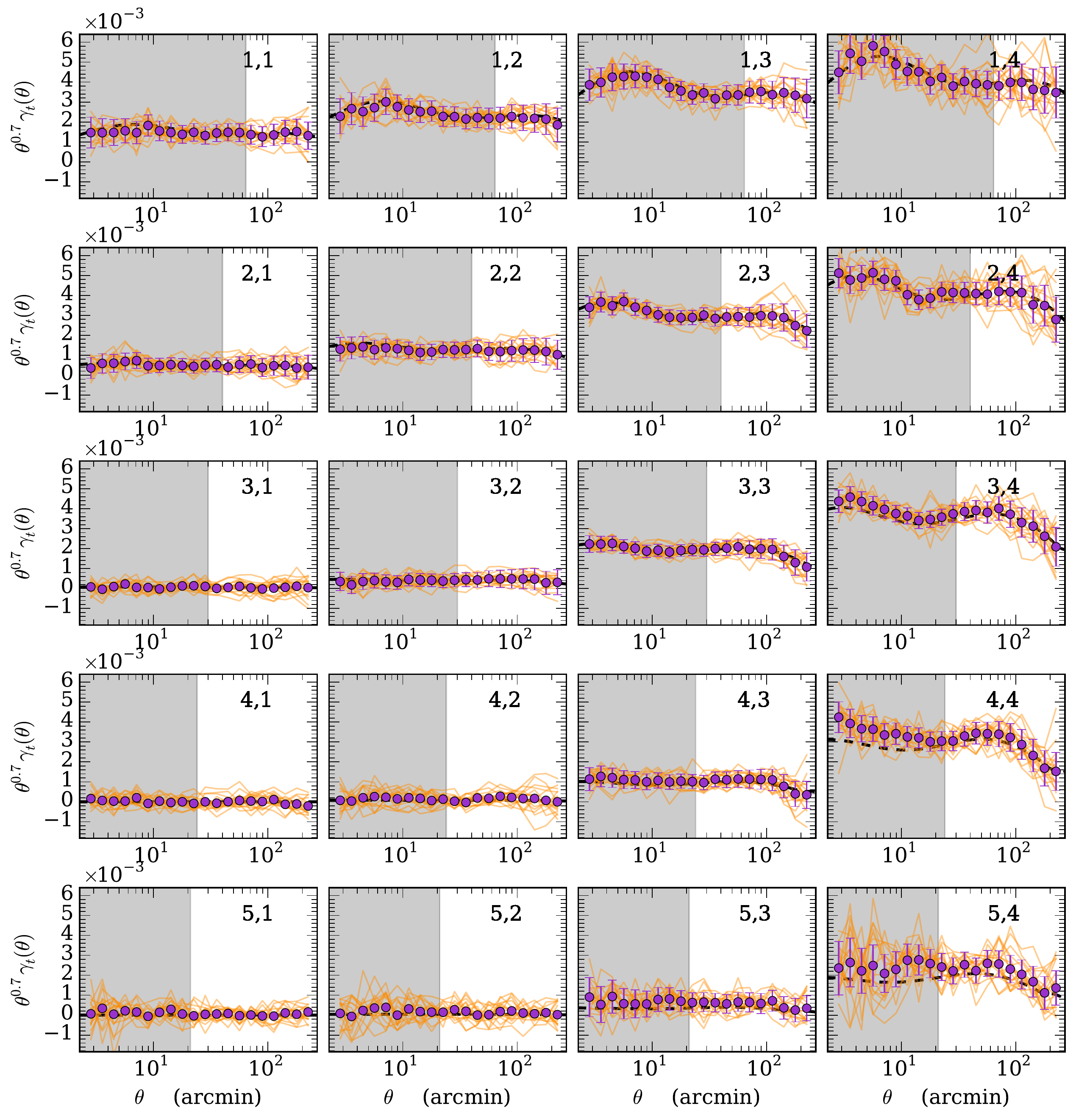}
\caption[]{Measurement of $\gamma_t(\theta)$ in the BCC simulations. Dark orchid data points are the mean across all 18 simulations, while orange lines indicate measurements from individual realizations. Errorbars indicate the expected uncertainty from a single DES Y1 realization. The line is the theoretical prediction assuming the true cosmology, with the best fit galaxy bias from the mean of all realizations. Grey shaded regions are excluded from the fiducial analysis.}
\label{fig:gammat}
\end{figure*}

\begin{figure*}
\includegraphics[width=14cm]{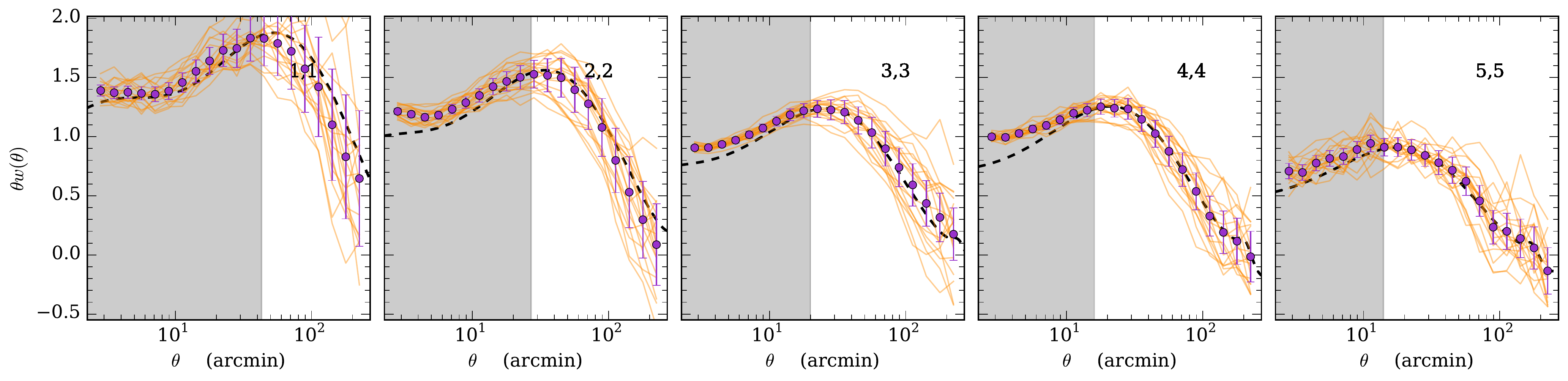}
\caption[]{Measurement of $\wtheta$ in the BCC simulations. Dark orchid data points are the mean across all 18 simulations, while orange lines indicate measurements from individual realizations. Errorbars indicate the expected uncertainty from a single DES Y1 realization. The line is the theoretical prediction assuming the true cosmology, with the best fit galaxy bias from the mean of all realizations. Grey shaded regions are excluded from the fiducial analysis.}
\label{fig:wtheta}
\end{figure*}

\begin{figure*}
\includegraphics[width=14cm]{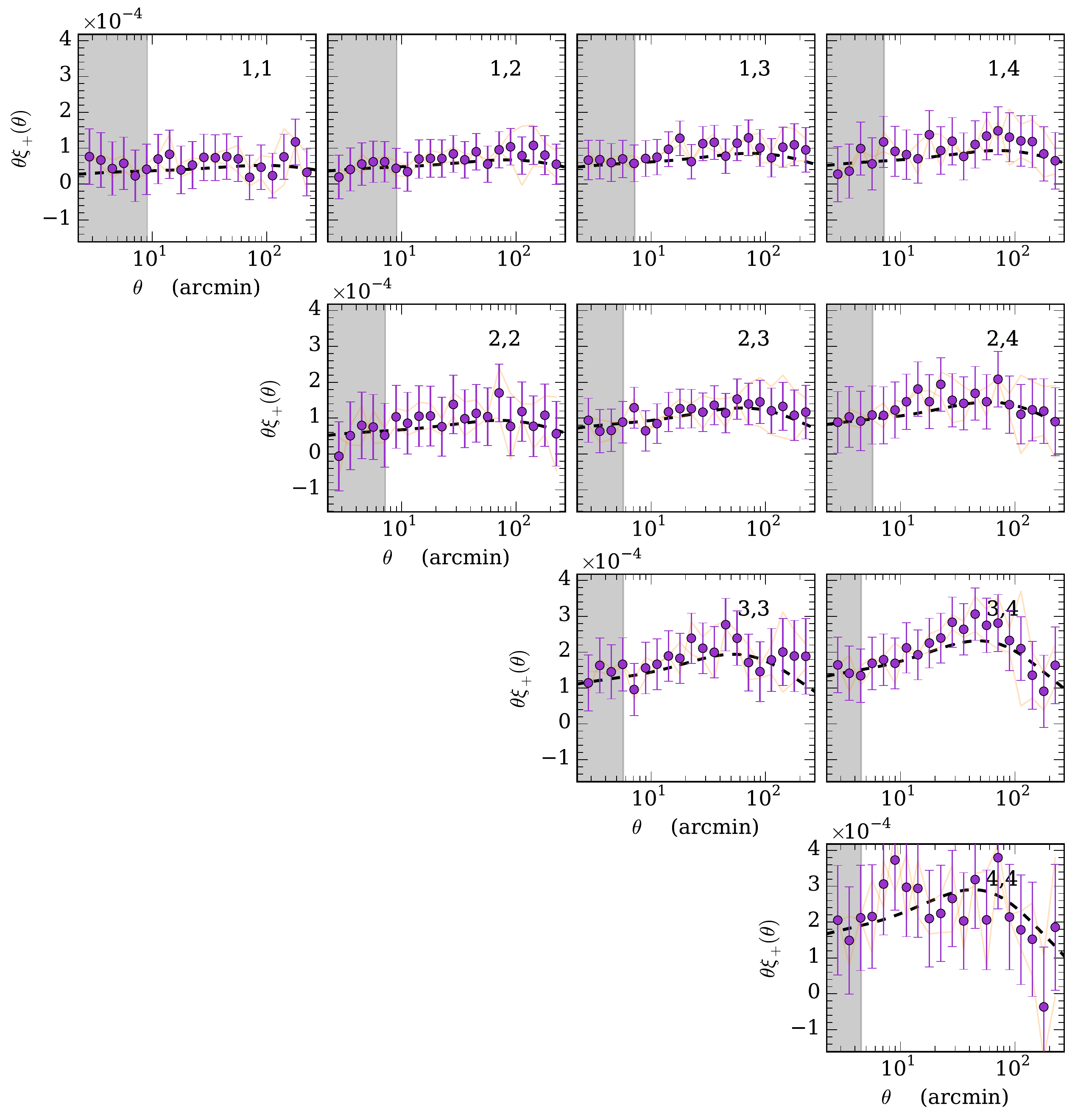}
\caption[]{Measurement of $\xip(\theta)$ in the MICE simulations. Dark orchid data points are the mean across the two DES Y1 realizations, while orange lines indicate measurements from individual realizations. Errorbars indicate the expected uncertainty from a single DES Y1 realization. The line is the theoretical prediction assuming the true cosmology. Grey shaded regions are excluded from the fiducial analysis.}
\label{fig:micexip}
\end{figure*}

\begin{figure*}
\includegraphics[width=14cm]{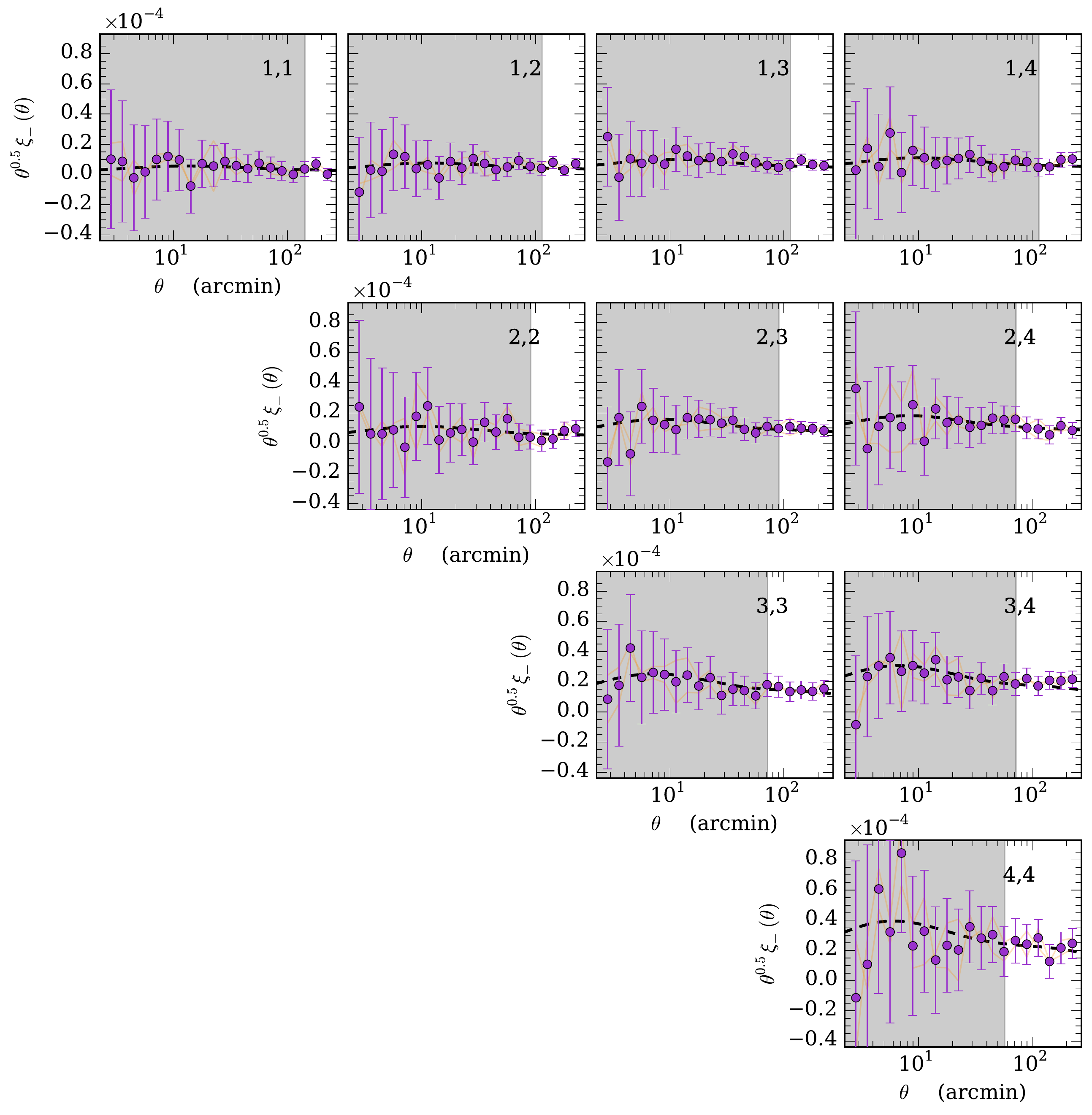}
\caption[]{Measurement of $\xim(\theta)$ in the MICE simulations. Dark orchid data points are the mean across the two DES Y1 realizations, while orange lines indicate measurements from individual realizations. Errorbars indicate the expected uncertainty from a single DES Y1 realization. The line is the theoretical prediction assuming the true cosmology. Grey shaded regions are excluded from the fiducial analysis.}
\label{fig:micexim}
\end{figure*}

\begin{figure*}
\includegraphics[width=14cm]{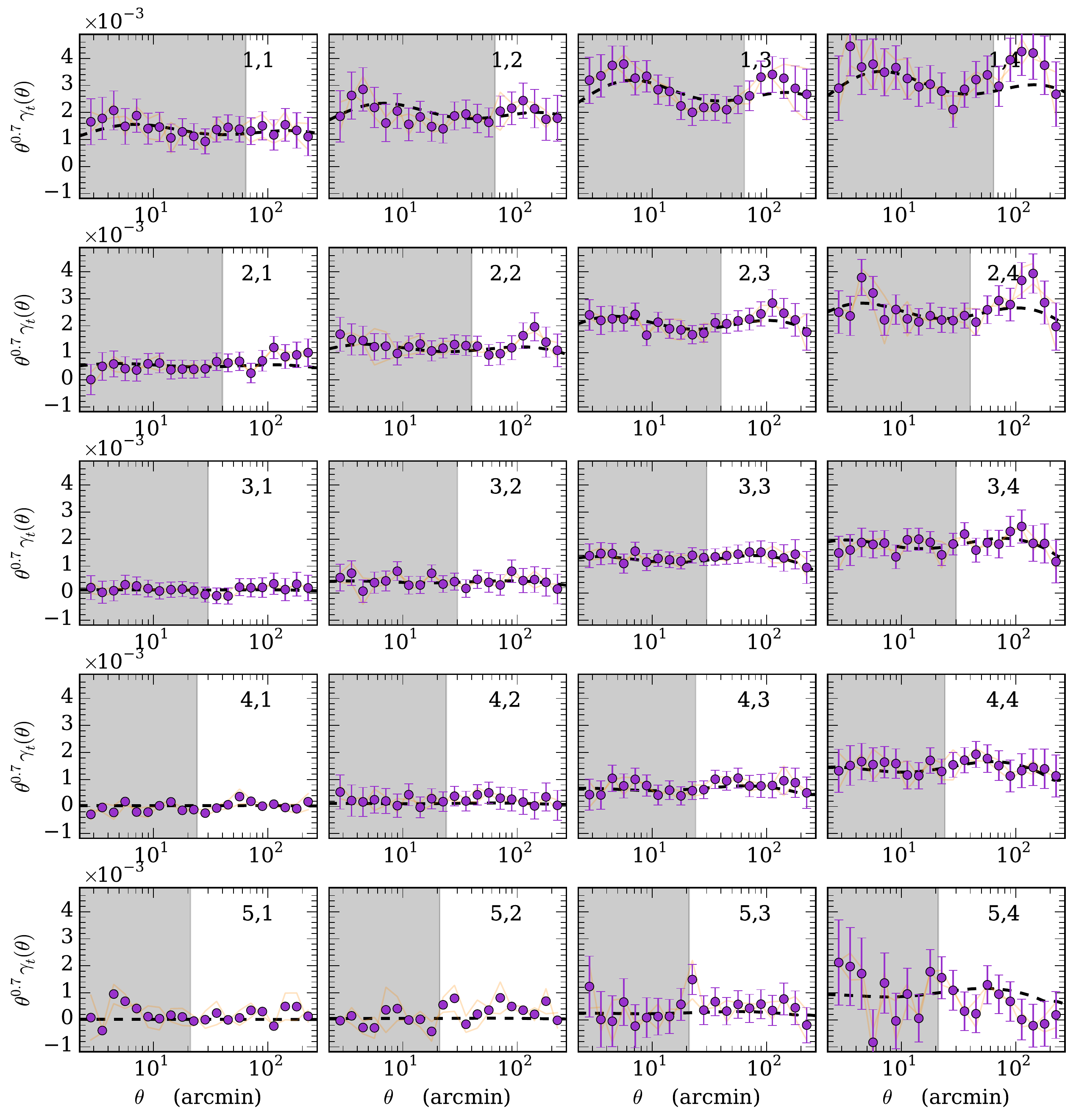}
\caption[]{Measurement of $\gamma_t(\theta)$ in the MICE simulations. Dark orchid data points are the mean across the two DES Y1 realizations, while orange lines indicate measurements from individual realizations. Errorbars indicate the expected uncertainty from a single DES Y1 realization. The line is the theoretical prediction assuming the true cosmology. Grey shaded regions are excluded from the fiducial analysis.}
\label{fig:micegammat}
\end{figure*}

\begin{figure*}
\includegraphics[width=14cm]{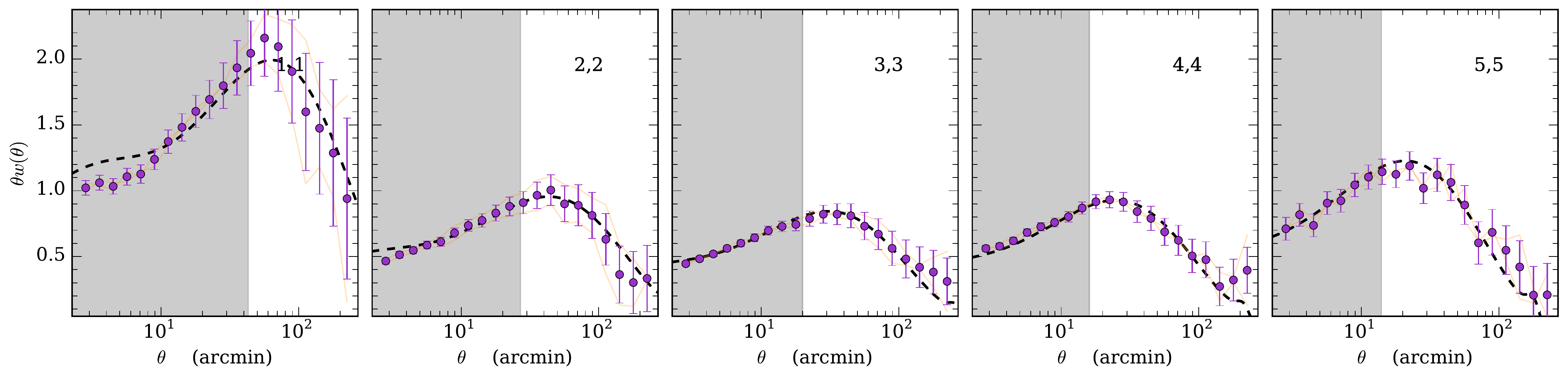}
\caption[]{Measurement of $\wtheta$ in the MICE simulations. Dark orchid data points are the mean across the two DES Y1 realizations, while orange lines indicate measurements from individual realizations. Errorbars indicate the expected uncertainty from a single DES Y1 realization. The line is the theoretical prediction assuming the true cosmology. Grey shaded regions are excluded from the fiducial analysis.}
\label{fig:micewtheta}
\end{figure*}

Figures~\ref{fig:xip}-\ref{fig:wtheta} show the two-point correlation function measurements from the BCC simulations. Figures~\ref{fig:micexip}-\ref{fig:micewtheta} show two-point correlation function measurements from the MICE simulations.

\section*{Affiliations}
$^{1}$ Center for Cosmology and Astro-Particle Physics, The Ohio State University, Columbus, OH 43210, USA\\
$^{2}$ Department of Physics, The Ohio State University, Columbus, OH 43210, USA\\
$^{3}$ Department of Physics, Stanford University, 382 Via Pueblo Mall, Stanford, CA 94305, USA\\
$^{4}$ Kavli Institute for Particle Astrophysics \& Cosmology, P. O. Box 2450, Stanford University, Stanford, CA 94305, USA\\
$^{5}$ SLAC National Accelerator Laboratory, Menlo Park, CA 94025, USA\\
$^{6}$ Institute of Physics, Laboratory of Astrophysics, \'Ecole Polytechnique F\'ed\'erale de Lausanne (EPFL), Observatoire de Sauverny, 1290 Versoix, Switzerland\\
$^{7}$ Institut d'Estudis Espacials de Catalunya (IEEC), 08193 Barcelona, Spain\\
$^{8}$ Institute of Space Sciences (ICE, CSIC),  Campus UAB, Carrer de Can Magrans, s/n,  08193 Barcelona, Spain\\
$^{9}$ Department of Physics and Astronomy, University of Pennsylvania, Philadelphia, PA 19104, USA\\
$^{10}$ Department of Astronomy/Steward Observatory, 933 North Cherry Avenue, Tucson, AZ 85721-0065, USA\\
$^{11}$ Jet Propulsion Laboratory, California Institute of Technology, 4800 Oak Grove Dr., Pasadena, CA 91109, USA\\
$^{12}$ Institute for Astronomy, University of Edinburgh, Edinburgh EH9 3HJ, UK\\
$^{13}$ Jodrell Bank Center for Astrophysics, School of Physics and Astronomy, University of Manchester, Oxford Road, Manchester, M13 9PL, UK\\
$^{14}$ Institut de F\'{\i}sica d'Altes Energies (IFAE), The Barcelona Institute of Science and Technology, Campus UAB, 08193 Bellaterra (Barcelona) Spain\\
$^{15}$ Fermi National Accelerator Laboratory, P. O. Box 500, Batavia, IL 60510, USA\\
$^{16}$ Department of Physics, Carnegie Mellon University, Pittsburgh, Pennsylvania 15312, USA\\
$^{17}$ Kavli Institute for Cosmological Physics, University of Chicago, Chicago, IL 60637, USA\\
$^{18}$ Department of Astronomy, University of Michigan, Ann Arbor, MI 48109, USA\\
$^{19}$ Department of Physics, University of Michigan, Ann Arbor, MI 48109, USA\\
$^{20}$ Cerro Tololo Inter-American Observatory, National Optical Astronomy Observatory, Casilla 603, La Serena, Chile\\
$^{21}$ Department of Physics \& Astronomy, University College London, Gower Street, London, WC1E 6BT, UK\\
$^{22}$ Department of Physics and Electronics, Rhodes University, PO Box 94, Grahamstown, 6140, South Africa\\
$^{23}$ Institute of Cosmology \& Gravitation, University of Portsmouth, Portsmouth, PO1 3FX, UK\\
$^{24}$ Laborat\'orio Interinstitucional de e-Astronomia - LIneA, Rua Gal. Jos\'e Cristino 77, Rio de Janeiro, RJ - 20921-400, Brazil\\
$^{25}$ Observat\'orio Nacional, Rua Gal. Jos\'e Cristino 77, Rio de Janeiro, RJ - 20921-400, Brazil\\
$^{26}$ Department of Astronomy, University of Illinois at Urbana-Champaign, 1002 W. Green Street, Urbana, IL 61801, USA\\
$^{27}$ National Center for Supercomputing Applications, 1205 West Clark St., Urbana, IL 61801, USA\\
$^{28}$ Centro de Investigaciones Energ\'eticas, Medioambientales y Tecnol\'ogicas (CIEMAT), Madrid, Spain\\
$^{29}$ Instituto de Fisica Teorica UAM/CSIC, Universidad Autonoma de Madrid, 28049 Madrid, Spain\\
$^{30}$ Department of Physics, ETH Zurich, Wolfgang-Pauli-Strasse 16, CH-8093 Zurich, Switzerland\\
$^{31}$ Santa Cruz Institute for Particle Physics, Santa Cruz, CA 95064, USA\\
$^{32}$ Max Planck Institute for Extraterrestrial Physics, Giessenbachstrasse, 85748 Garching, Germany\\
$^{33}$ Universit\"ats-Sternwarte, Fakult\"at f\"ur Physik, Ludwig-Maximilians Universit\"at M\"unchen, Scheinerstr. 1, 81679 M\"unchen, Germany\\
$^{34}$ Harvard-Smithsonian Center for Astrophysics, Cambridge, MA 02138, USA\\
$^{35}$ Australian Astronomical Observatory, North Ryde, NSW 2113, Australia\\
$^{36}$ Departamento de F\'isica Matem\'atica, Instituto de F\'isica, Universidade de S\~ao Paulo, CP 66318, S\~ao Paulo, SP, 05314-970, Brazil\\
$^{37}$ George P. and Cynthia Woods Mitchell Institute for Fundamental Physics and Astronomy, and Department of Physics and Astronomy, Texas A\&M University, College Station, TX 77843,  USA\\
$^{38}$ Instituci\'o Catalana de Recerca i Estudis Avan\c{c}ats, E-08010 Barcelona, Spain\\
$^{39}$ School of Physics and Astronomy, University of Southampton,  Southampton, SO17 1BJ, UK\\
$^{40}$ Brandeis University, Physics Department, 415 South Street, Waltham MA 02453\\
$^{41}$ Instituto de F\'isica Gleb Wataghin, Universidade Estadual de Campinas, 13083-859, Campinas, SP, Brazil\\
$^{42}$ Computer Science and Mathematics Division, Oak Ridge National Laboratory, Oak Ridge, TN 37831\\
$^{43}$ Excellence Cluster Universe, Boltzmannstr.\ 2, 85748 Garching, Germany\\

\bsp	
\label{lastpage}
\end{document}